\documentclass[twocolumn,showpacs,aps,prd,amsmath,amssymb,nobibnotes,floatfix]{revtex4}
\usepackage{hyperref}

\usepackage{mathtools}
\usepackage{graphicx,psfrag}
\usepackage{amsmath,amsfonts,amssymb,amsthm}
\usepackage{hyperref}
\usepackage{times}
\usepackage{color}

\begin{document}

\title{Magnetorotational collapse of supermassive stars: \\ Black hole
  formation, gravitational waves, and jets}
\author{Lunan Sun$^1$}
\author{Vasileios Paschalidis$^2$}
\author{Milton Ruiz$^1$}
\author{Stuart L. Shapiro$^{1,3}$}
\affiliation{$^1$
  Department of Physics, University of Illinois at Urbana-Champaign, Urbana, Illinois 61801, USA\\
  $^{2}$Department of Physics, Princeton University, Princeton, New Jersey 08544, USA \\
  $^{3}$Department of Astronomy \& NCSA, University of Illinois at Urbana-Champaign, Urbana, Illinois 61801, USA
}

\begin{abstract}
We perform magnetohydrodynamic simulations in full general relativity
of uniformly rotating stars that are marginally unstable to collapse.
These simulations model the direct collapse of supermassive stars (SMSs) to
seed black holes that can grow to become the supermassive black holes
at the centers of quasars and active galactic nuclei. They also
crudely model the collapse of massive Population III stars to black
holes, which could power a fraction of distant, long gamma-ray
bursts. The initial stellar models we adopt are $\Gamma = 4/3$
polytropes 
initially with a dynamically unimportant dipole magnetic field. We
treat initial magnetic-field configurations either confined to the
stellar interior or extending out from the stellar interior into the
exterior. We find that the black hole formed following collapse has
mass $M_{\rm BH} \simeq 0.9M$ (where $M$ is the mass of the initial
star) and dimensionless spin parameter $a_{\rm BH}/M_{\rm BH}\simeq
0.7$. A massive, hot, magnetized torus surrounds the remnant black
hole.  At $\Delta t\sim 400-550M\approx 2000-2700(M/10^6M_\odot)$s
following the gravitational wave peak amplitude, an incipient jet is
launched. The disk lifetime is $\Delta t\sim 10^5(M/10^6M_\odot)$s,
and the outgoing Poynting luminosity is $L_{\rm EM}\sim
10^{51-52}$ ergs/s. If $\gtrsim 1\%-10\%$ of this power is converted
into gamma rays, Swift and Fermi could potentially
detect these events out to large redshifts $z\sim 20$. Thus,
SMSs could be sources of
ultra-long gamma-ray bursts (ULGRBs) and  massive Population III stars
could be the progenitors that power a fraction of the  long GRBs observed
at redshift $z \sim 5-8$. Gravitational waves are copiously emitted
during the collapse and peak at $\sim 15(10^6 M_{\odot}/M)\rm mHz$
[$\sim 0.15(10^4 M_{\odot}/M)\rm Hz$], i.e., in the LISA (DECIGO/BBO)
band; optimally oriented SMSs could be detectable by LISA
(DECIGO/BBO) at $z \lesssim 3$ ($z \lesssim 11$). 
Hence, $10^4 M_{\odot}$ SMSs collapsing at $z\sim
10$ are promising multimessenger sources of coincident gravitational
and electromagnetic waves.
\end{abstract}
\maketitle
\bigskip

\section{Introduction}
\label{section:Introduction}
Accreting supermassive black holes (BHs) are believed to be the
engines that power quasars and active galactic nuclei (AGNs).
Supermassive BHs (SMBHs) with mass
$\gtrsim 10^9 M_\odot$ are thought to reside in the centers of
quasars that have been detected as far as redshift $z \sim 7$
\cite{MorWarVen11} (see~\cite{Fan06} for a review of
high-redshift quasars). The detection of $10^9M_\odot$  SMBHs
at such high redshifts poses a major theoretical problem
(see~\cite{Hai12,Latif:2016qau,Smith2017arXiv170303083S} for recent
reviews): how could BHs as massive as a few billion times the mass of
our Sun form so early in the course of the evolution of our Universe?

It has been suggested that first generation---Population III (Pop
III)---stars could collapse and form seed BHs at large cosmological
redshifts, which later could grow through accretion to become
SMBHs~\cite{MadRee01,Heger:2002by}. This is possible because Pop III
stars with masses in the range $25-140 M_\odot$ and $> 260 M_\odot$
can undergo collapse to a BH~\cite{HegWoo02} at the end of their
lives. In turn, a $\sim 100M_\odot$ seed BH that accretes at the
Eddington limit with $\lesssim 10\%$ efficiency can grow to $M_{\rm
  BH}\gtrsim 10^9 M_\odot$ by $z\sim 7$, if the onset of accretion is
at $z\gtrsim 20$~\cite{Sha05,Alvarez:2008vw}.  Thus, accretion onto
BHs formed following the collapse of Pop III stars seems a viable
explanation for the origin of SMBHs by $z\sim 7$. However, this
scenario has a drawback because it has been argued that BHs cannot
grow at the Eddington limit over their entire history. In particular,
photoionization, heating and radiation pressure combine to modify the
accretion flow and may reduce it to $\sim 1/3$ of the
Eddington-limited rate~\cite{Alvarez:2008vw,
  Milosavljevic:2008vx}. One way to reconcile it is to combine the
accretion with mergers of seed BHs into their gaseous center in a cold
dark matter (CDM) model (see, e.g.  Refs. \cite{VolHaaMad02, Hai04,
  Sha05}). Simulations on assembling SMBHs using Monte Carlo merger
tree methods provide possible sub-Eddington growth models for Pop III
progenitors (see, e.g.
Refs. \cite{Tanaka2014CQGra..31x4005T,Volonteri:2003kr}).

An alternative scenario explaining the origin of SMBHs is provided by
the direct collapse of stars with masses $M\gtrsim
10^{4-5}M_\odot$~\cite{Ree84,BegVolRee06,Beg10} (see also
~\cite{Loeb:1994wv,Oh:2001ex,Bromm:2002hb,Koushiappas:2003zn,
  Lodato:2006hw,Shapiro:2003ua,Sha03}).
These so-called supermassive stars (SMSs) could form in metal-, dust-, and
$H_2$-poor halos, where fragmentation and formation of smaller stars
with masses $ < 100M_\odot$ could be suppressed (see, e.g. Refs
\cite{Regan14,Ress84,Gnedin:2001ey}).

Recent stellar evolution calculations suggest that SMSs can form, if
rapid mass accretion ($\dot{M}\gtrsim 0.1M_\odot/\text{yr}$) takes
place~\cite{Hosokawa2013},
and that the inner core can become unstable against collapse to a BH
once the stellar mass reaches $M = \mbox{few} \times
10^5M_\odot$. Even though the initial super-Eddington growth of a
black hole formed by SMS direct collapse could stop when the BH mass
reaches $10^3 - 10^4M_\odot$, it has been argued that the mass could
increase to $\sim 10^6M_\odot$ by $z \sim 10$~\cite{BegVolRee06}. These
more massive seed BHs could grow through accretion at sub-Eddington
rates (though not much less than 10\% -- 20\% of the Eddington accretion
rate~\cite{Tanaka2014CQGra..31x4005T}) to form the observed SMBHs, and
would require such rapid accretion over a shorter time window than the
seed BHs that may form in the collapse of Pop III stars.

However, the issue of fragmentation
inside the halos, where SMSs may form, is not entirely
resolved~\cite{Hai12,TanLiHai13,InaHai14,VisHaiBry14}. Nevertheless,
recent calculations suggest that fragmentation can be suppressed
either by turbulence~\cite{MayFiaBon14} (see also~\cite{BegVolRee06})
or through the dissociation of molecular
hydrogen~\cite{Fernandez:2014wia} via shocks or due to a Lyman-Werner
radiation background (see, e.g., Ref. ~\cite{VisHaiBry14} and references
therein). In addition, a recent study of baryon streaming on large scales with
respect to the dark matter indicates an alternative mechanism for delaying
Pop III and massive star formation ~\cite{SchRegGlo17}. Therefore, if
fragmentation is suppressed, the SMS direct-collapse framework appears to
provide a reasonable solution to the presence of $\gtrsim 10^9M_\odot$
SMBHs by $z\sim 7$. However, any model that explains the presence of
$10^9M_\odot$ SMBHs by $z\sim 7$ should also be able to explain the mass
distribution of SMBHs, and this does not seem to be an easy task. For
example, success in explaining the number of $\sim 10^9M_\odot$ SMBHs could
result in an overproduction of smaller mass BHs~\cite{Tanaka:2008bv}. One
possibility is raised by a recent semianalytic model assuming warm dark
matter (WDM) cosmology~\cite{DayChoPac17}, in which the BH density increases
by direct collapse from z=17.5 to z=8, and structure formation is such that
``pristine" halos with virial temperatures $T > 10^4 K$ form up to z=5. This
implies that environments favorable for forming SMSs that can undergo direct
collapse could appear even at z=5, peaking at z=8. These results provide
a promising opportunity for multimessenger observations.

Despite the progress in understanding the astrophysics of SMSs,
much work is left to be done, both theoretically and
observationally. For example, while
conditions allowing the formation and direct collapse of SMSs may be
present at cosmological redshifts $z \gtrsim10$
\cite{Smith2017arXiv170303083S}, indirect observational evidence
for the existence of SMSs at high redshifts appears
controversial~\cite{Smith2017arXiv170303083S}. This fact may change
with future telescopes that will probe the high-redshift
Universe~\cite{Smith2017arXiv170303083S}.  Moreover, it remains an
open question when and where in the Universe conditions favorable for
forming SMSs are found, and as a result, rates of formation and
collapse of SMSs as a function of $z$ are currently uncertain, as are
the processes that may limit the growth of SMS-formed seed
BHs~\cite{Tanaka:2008bv}.

It is not inconceivable that SMSs could form even at $z
\lesssim 10$ in the right environment.
If that is the case, collapsing SMSs could generate detectable transient
gravitational wave (GW) and electromagnetic (EM) signatures. The
multimessenger signatures from the direct collapse and subsequent
hyper-accretion phase of SMSs have not been explored to a great extent.
To facilitate the interpretation of future transient GW and EM observations,
a theoretical effort targeted at predicting the multimessenger
signatures of such collapsing and hyperaccreting SMSs is required. It
could be that a collapsing SMS may power an ultra-long gamma-ray burst
(ULGRB). Such a burst could be observable even at very large redshifts.
If the SMS has the right mass the GW burst generated during
the collapse, black hole formation and ringdown could be detectable by
future space-based GW observatories. Detection of such multimessenger
signals would provide smoking-gun evidence for the SMS direct-collapse
origin of seed SMBHs.

As SMSs may form by the accretion of magnetized, collapsing primordial
gas clouds~(see~\cite{BanJed04,Silk:2006ja,SchBanSur10,SurFedSch12,
  TurOisAbe12, MacDoi13}), it is likely that they are
magnetized and spinning. Radiative cooling accompanied by mass loss may
induce quasistatic contraction that spins up the star to  near the
mass-shedding limit on a secular time scale ~\cite{BauSha99}. The presence
of magnetic-induced turbulent viscosity will damp differential rotation
and drive the star to uniform rotation. Upon reaching the general
relativistic onset of radial instability, the star will
collapse on a dynamical time scale and, eventually, form a spinning BH
\cite{ZelNov71, BauSha99}. All of the above features motivate studies in full
general relativity of the magnetorotational collapse of SMSs.

Recent GR hydrodynamic calculations~\cite{ShiUchSek16,
  Shibata:2016vzw} suggest that the equation of state (EOS) of a
rigidly rotating SMS core, marginally unstable to collapse, may be
better approximated by a $\Gamma\approx 1.335\gtrsim 4/3$ polytrope.
However, since SMSs are convective and their EOS is dominated by
thermal radiation pressure, they can be well approximated by simple
$\Gamma = 4/3$ polytropes. Multiple collapse simulations of
$\Gamma=4/3$ polytropes have been performed in the past. Apart from
the simplicity of this EOS, another advantage of such polytropes is
that they can model not only SMSs, but also
massive Pop III stars, albeit crudely, that  also
collapse and form BHs. Such collapsing massive Pop III stars could
potentially power observable, transient EM signals. For example, while
long gamma-ray bursts (lGRBs) are thought to originate in the
core collapse of massive, low-metallicity stars, the recent discovery
of Swift's Burst Alert Telescope (BAT) sources at cosmological redshifts $z
\sim 5.3 - 8.0$ (see, e.g. Refs. ~\cite{swiftGRB140304A, swiftGRB090423}),
raises the exciting possibility that some of these explosions may
originate in the collapse of massive, metal-free (Pop III) stars. This
is because the star formation density of Pop III stars is predicted to
peak at z $\sim$ $5 - 8$ (see, e.g. Refs. ~\cite{TorFerSch07,JohVecKho13}),
which is consistent with recent observations supporting the discovery
of a population of Pop III stars at redshift z $\sim$ 6.5
\cite{SobMatDar15}. 

GR simulations of the collapse of marginally unstable, nonrotating
SMSs were first performed
in~\cite{ShaTeu79} adopting an initial $\Gamma \approx 4/3$ polytrope in
spherical symmetry, where it was concluded that $90\%$ of the initial rest
mass would fall into the BH in a time $\lesssim 30 M$ after its
appearance. Subsequently, axisymmetric simulations of rotating SMS collapse
were performed in~\cite{ShiSha02,LiuShaSte07}. The GR
hydrodynamic calculations of marginally unstable, uniformly rotating SMSs
that spin at
the mass-shedding limit in~\cite{ShiSha02,LiuShaSte07} found that
about $90\% -- 95\%$ of initial stellar mass forms a spinning BH with spin
parameter $a_{\rm BH}/M_{\rm BH} \sim 0.7-0.75$. They also found
that the remnant BH is surrounded by a massive, hot accretion
torus.
An analytic treatment~\cite{ShaShi02} was able to corroborate
many of these results and verify that the final, nondimensional
BH spin and disk parameters were independent of the progenitor mass.
In the absence of initial nonaxisymmetic perturbations, differential
rotation does not induce any significant changes in the final
BH-accretion disk configuration~\cite{Sai04, SaiHaw09}.

Axisymmetric GR magnetohydrodynamic (GRMHD) calculations of an
unstable $\Gamma = 4/3$ polytrope, rotating uniformly at the
mass-shedding limit were performed in ~\cite{LiuShaSte07}. The authors
seeded the initial star with a poloidal magnetic field confined to its
interior, and showed that the final configuration 
consisted of a central BH surrounded by a massive, hot accretion
torus. The emergence of a collimated magnetic field above the BH poles
was reported, but the evolution could not be followed  too long
after BH formation.  The authors speculated that the system might
eventually launch a relativistic jet.

The collapse of SMSs is also a source of GWs ~\cite{LiuShaSte07,ShiSekUch16}.  In \cite{ShiSekUch16} it was
found that the GW signal produced by the collapse of a $6.3\times 10^5
M_\odot$ SMS at redshift $z=3$ peaks at frequency $\sim 5{\rm mHz}$,
and could be detectable by a LISA-like detector. GRMHD simulations
in~\cite{LiuShaSte07} showed that magnetic fields can induce episodic radial
oscillations in the accretion disk, which may generate long-wavelength
GWs that could be detectable at $z \sim 5$ for $M_{\text{SMS}}\gtrsim
10^4 M_\odot$.

In this work we extend previous GR simulations of collapsing massive
stars in several ways: (a) we lift the assumption of axisymmetry and
perform simulations in 3+1 dimensions, (b) we introduce magnetic fields
that are initially dynamically unimportant and are either confined to
the stellar interior or extend out from the stellar interior into the
exterior, (c) we follow the post-BH formation evolution for much longer
times than previous works through jet launching.
We adopt the same initial stellar equilibrium model as
in~\cite{LiuShaSte07}. Following collapse, and once the remnant
BH-disk system has settled to a quasistationary state, we find that
the mass and dimensionless spin parameter of the BH are consistent
with those reported in~\cite{ShiSha02,LiuShaSte07}.  We find that
about $\Delta t \approx 400-550M\approx 2000-2700(M/10^6M_\odot)$s
after BH formation, our magnetized configurations launch a strongly
magnetized, collimated, and mildly relativistic outflow---an
incipient jet (cf.  \cite{PasRuiSha15,RuiLanPas16}). We estimate that
these jets could power gamma-ray bursts that may be detectable by Swift and Fermi.
For SMSs with masses of $10^6M_\odot$, the resulting GWs peak in the LISA band and optimally oriented sources could be detectable
at $z\lesssim 3$; however for SMSs with masses of $10^4M_\odot$ the GWs
peak in the (Decihertz Interferometer Gravitational Wave Observatory/Big Bang Observer (DECIGO/BBO) band, and optimally orientated sources
could be detectable by DECIGO at $z\lesssim 8$, and by BBO at $z\lesssim 11$.

The paper is organized as follows. In Sec.~\ref{sec:num_setup} we
present a detailed description of the initial data we adopt 
and describe our numerical methods and the diagnostics we use to monitor
our calculations. In Sec.~\ref{section:Numerical Results} we present
our results, and in Sec.~\ref{section:Observational Prospects} we
discuss their implications for the detection of GW and EM signals. We
conclude in Sec.~\ref{section:Summary and Conclusions} with a brief
summary
and a discussion of future work. Unless
otherwise stated, we adopt geometrized units ($G=c=1$) throughout.

\begin{figure*}[t]
\centering
\includegraphics[width=0.49\textwidth]{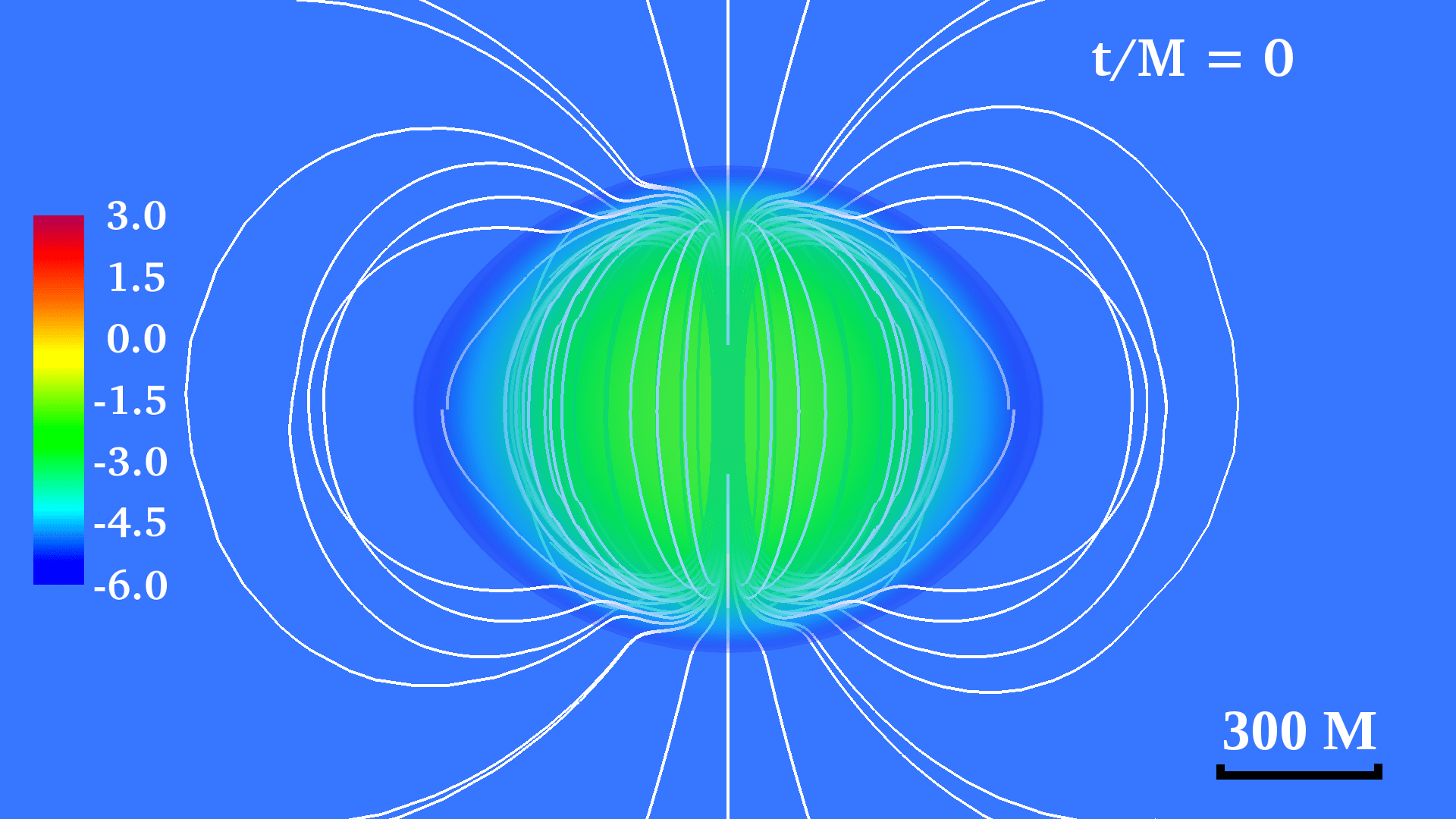}
\includegraphics[width=0.49\textwidth]{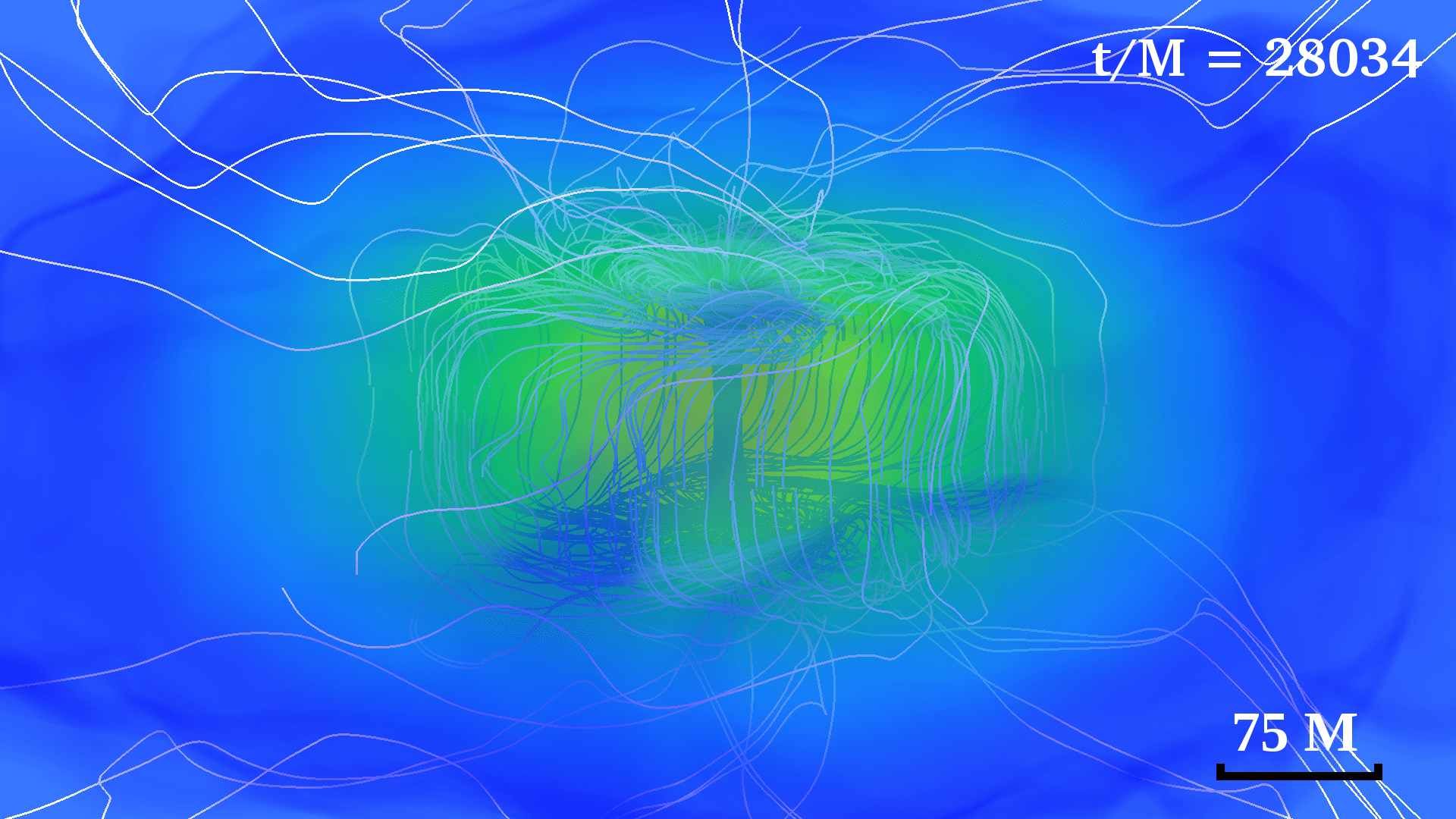}
\includegraphics[width=0.49\textwidth]{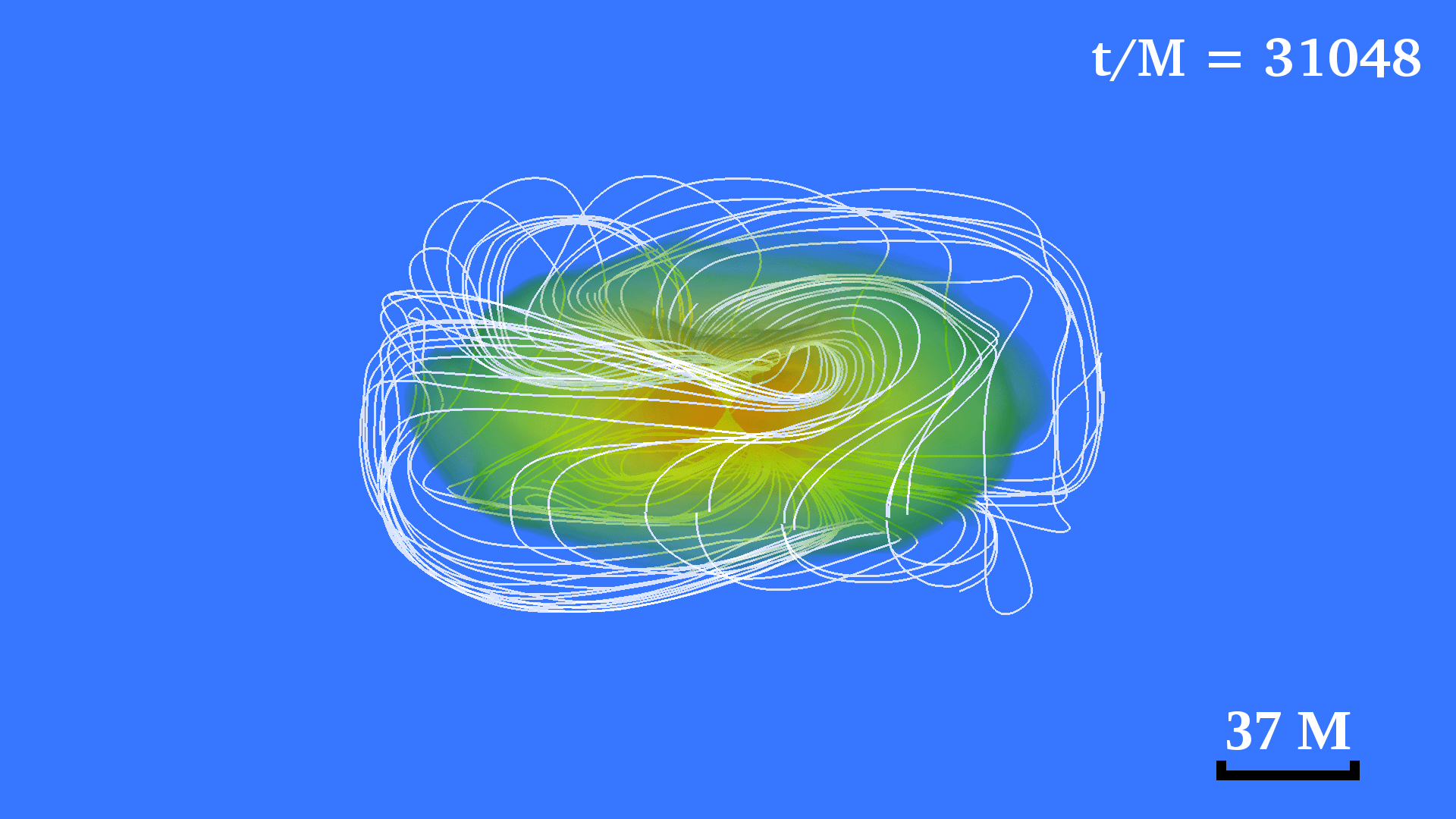}
\includegraphics[width=0.49\textwidth]{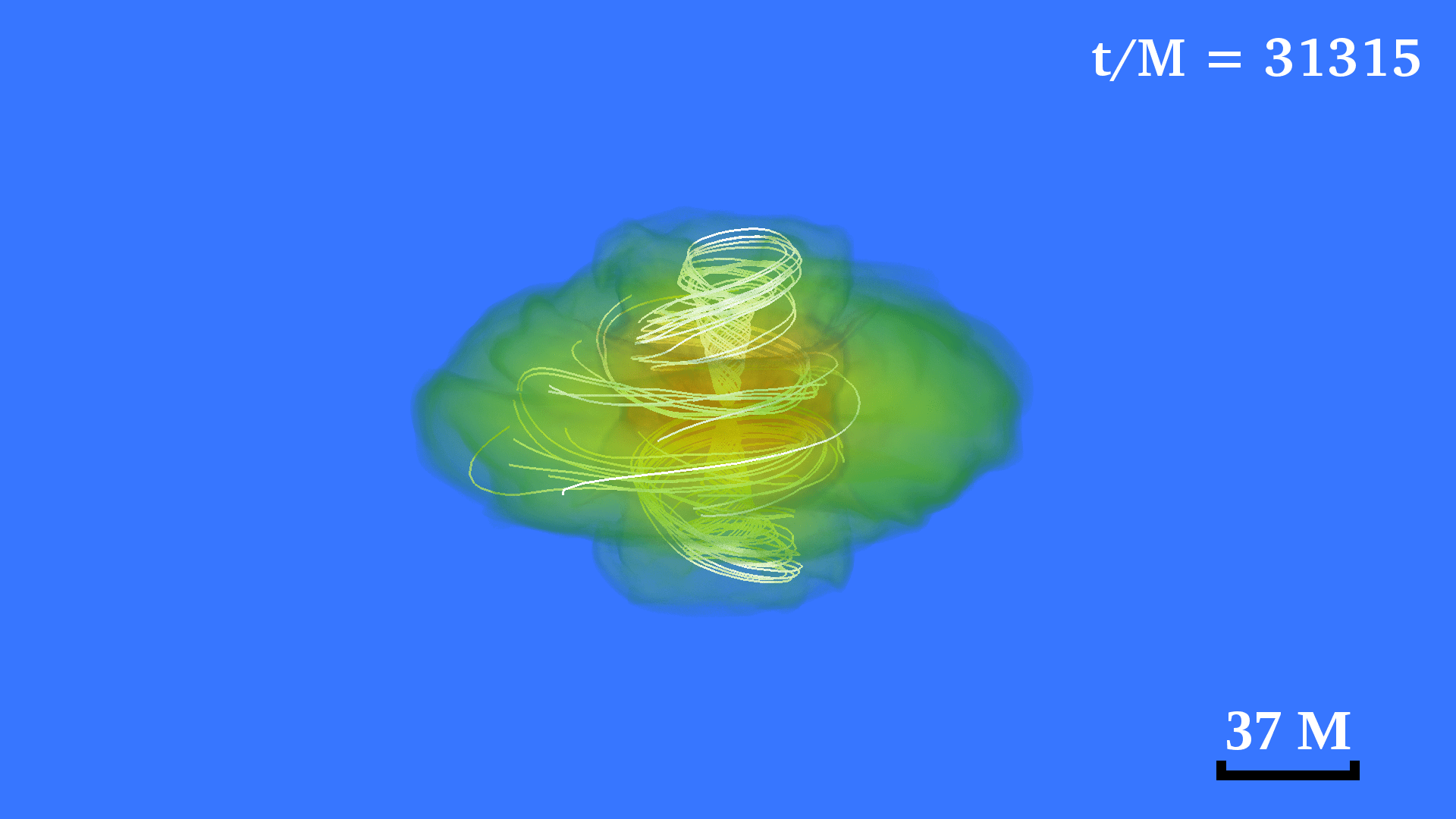}
\includegraphics[width=0.49\textwidth]{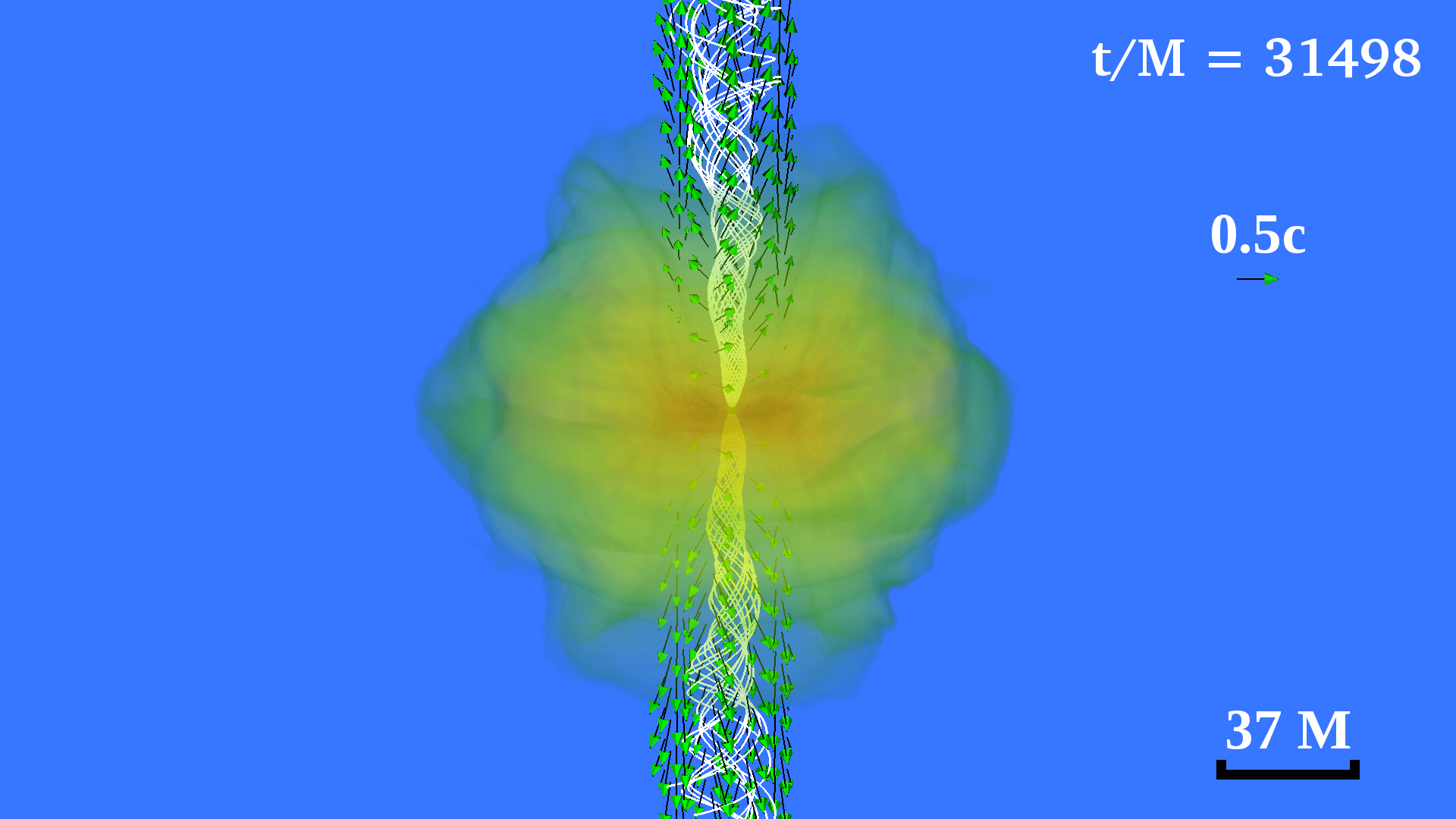}
\includegraphics[width=0.49\textwidth]{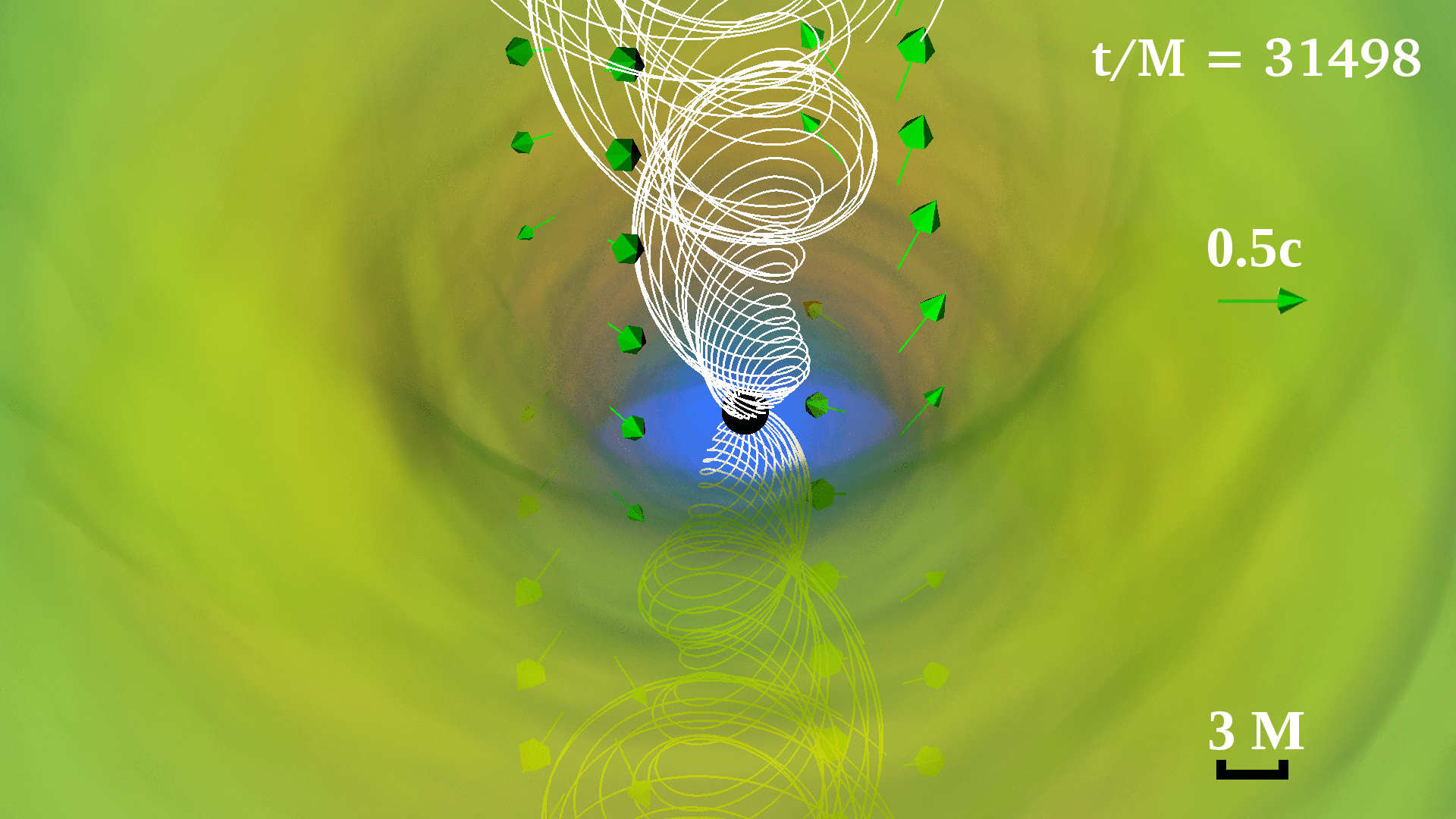}
\caption{\label{fig:Evolution} Volume rendering of the rest-mass
  density normalized to its initial maximum value
  $\rho_{0,\text{max}}=9.9\times10^{-2}(M/10^6M_{\odot})^{-2}\,{\rm g}\, {\rm
    cm}^{-3}$ (log scale) at select times for the $\rm
  S_{\rm{\tiny{Int+Ext}}}$ case.  Solid lines indicate the magnetic-field lines, and arrows show plasma velocities with length
  proportional to their magnitude. The bottom left panel displays
  the collimated, helical magnetic field and outgoing plasma, whose zoomed-in
  view near the horizon is shown in the bottom right panel. Here
  $M=4.9 (M/10^6M_\odot){\rm s}=1.47\times 10^6(M/10^6M_\odot)\rm
  km$. }
\end{figure*}

\section{Methods}
\label{sec:num_setup}

In this section we describe in detail our initial data, the
numerical method and the grid structure we employ for solving
the Einstein equations coupled to the equations of ideal
magnetohydrodynamics in a dynamical, curved spacetime. We
also summarize the diagnostics we adopt to monitor the
simulations.

\subsection{Initial data}
To model a collapsing SMS, and also to crudely model the collapse of a
Pop III star, we start with a marginally unstable $\Gamma=4/3$
polytrope that is uniformly rotating at the mass-shedding limit. The
rotating polytropic star is built with the code of~\cite{CooShaTeu92,
  CooShaTeu94}. We employ dimensionless (barred) variables in which,
for instance, the radius $R$, mass $M$ and density $\rho$ are scaled
as follows~\cite{Baumgarte10}
\begin{align}
\bar{R}=&\kappa^{-n/2}\,R,\qquad
\bar{M}=&\kappa^{-n/2}\,M,\qquad
\bar{\rho}= \kappa^{n}\,\rho\,,
\end{align}
where $n=1/(\Gamma-1)$ is the polytropic index.  Our calculations
scale with the polytropic constant $\kappa$. The polytropic model we
adopt has the same initial properties as the one
in~\cite{LiuShaSte07}, and it is characterized by the following
parameters: ADM mass $\bar{M}_{\rm ADM}=4.572$, central rest-mass
density $\bar{\rho}_{0,c} = 7.7\times 10^{-9}$, dimensionless angular
momentum $J/M^2=0.96$, and ratio of kinetic to gravitational-binding
energy $T/|W|=0.009$.  The equatorial radius of the star is
$R_{\rm eq}=626M_{\rm ADM}\approx 9.25\times
10^6\,({M_{\rm ADM}}/{10^6M_{\odot}})$km. This model is marginally
unstable to collapse.

\begin{figure*}[t]
\includegraphics[width=0.49\textwidth]{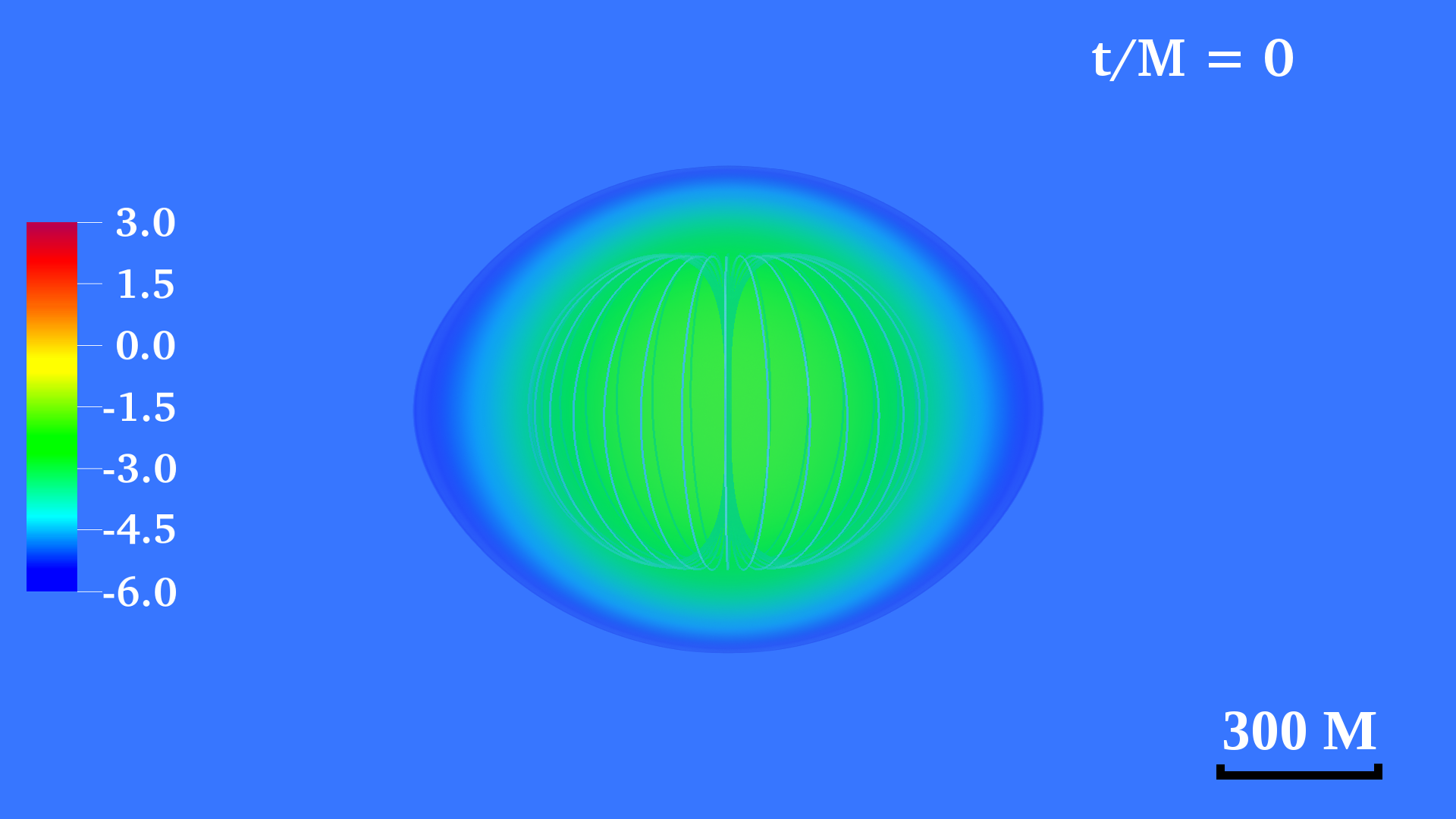}
\includegraphics[width=0.49\textwidth]{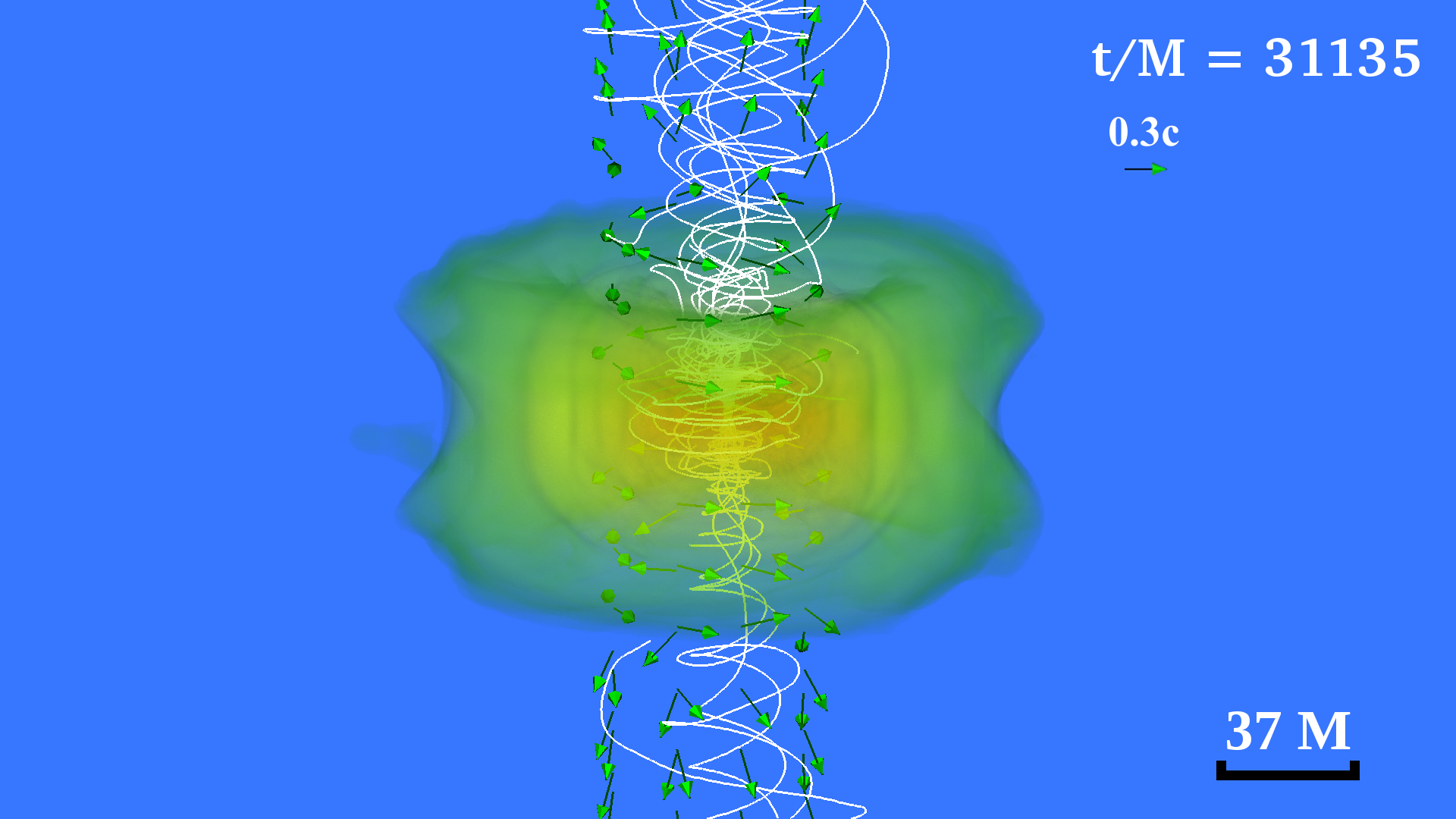}
\includegraphics[width=0.49\textwidth]{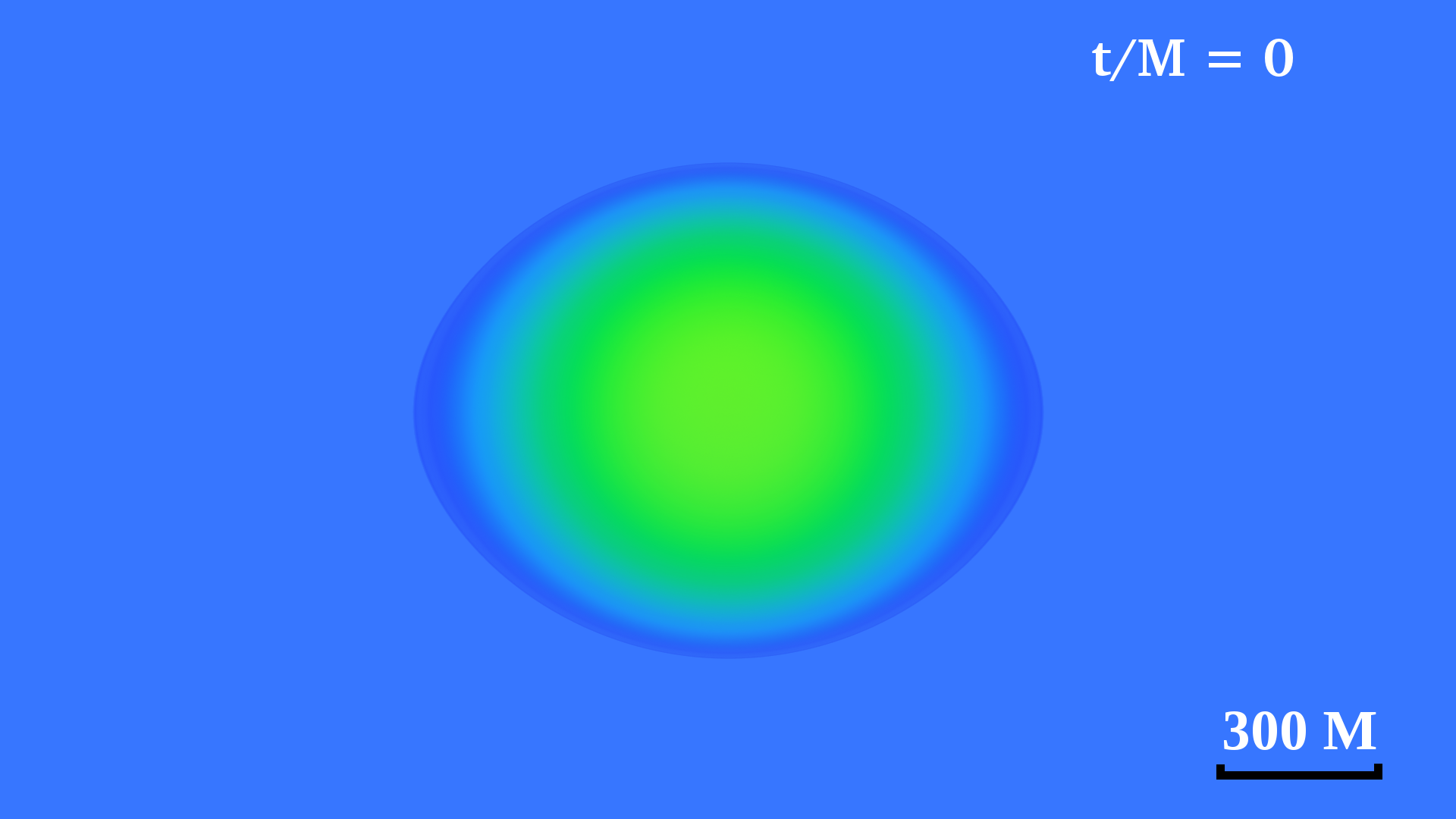}
\includegraphics[width=0.49\textwidth]{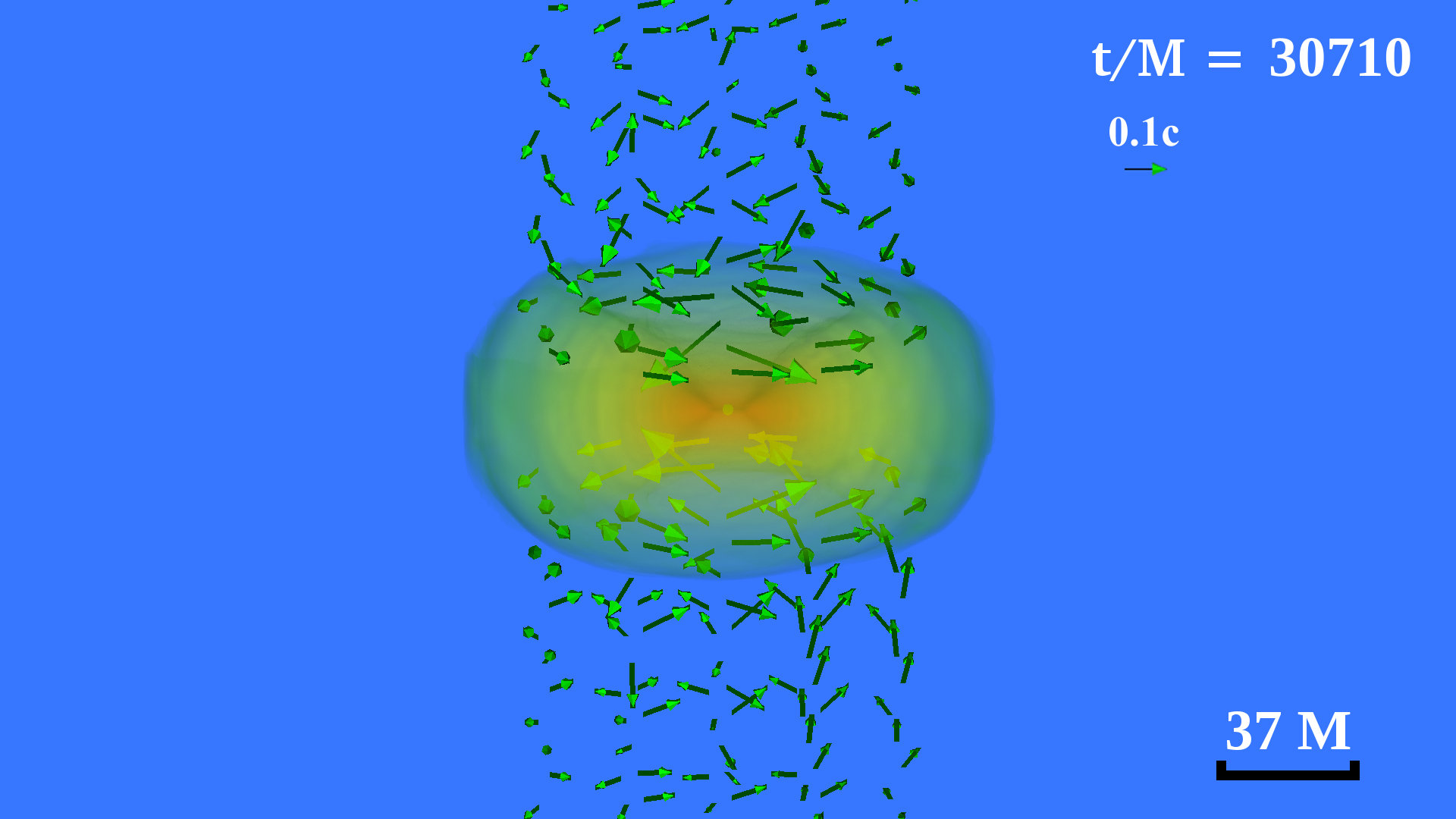}
  \caption{\label{fig:initial} Volume rendering of the rest-mass
    density normalized to its initial maximum value (log scale) for
    the case $\rm S_{\rm{\tiny{Int}}}$ (top row) and the case $\rm
    S_{\rm{\tiny{Hydro}}}$ (bottom row). Initial and final
    configurations for these two cases are shown in the left and right
    panels, respectively. Solid lines indicate the magnetic-field
    lines, and arrows show plasma velocities with length proportional
    to their magnitude.  Here $M=4.9(M/10^6M_\odot){\rm s}=1.47\times
    10^6(M/10^6M_\odot)\rm km$. }
\end{figure*}

We consider three different initial scenarios as follows: 
\begin{itemize}
\item Case $\rm S_{\rm{\tiny{Int+Ext}}}$: Magnetized configuration in
  which the initial equilibrium star is seeded with a dipole-like
  magnetic field which extends from the stellar interior into the
  exterior (see top left panel in Fig.  \ref{fig:Evolution}),
\item Case $\rm S_{\rm{\tiny{Int}}}$: Magnetized configuration in
  which the initial equilibrium star is seeded with a poloidal
  magnetic field confined to the stellar interior (see top left panel
  in Fig.~\ref{fig:initial}),
  \item Case $\rm S_{\rm{\tiny{Hydro}}}$: Purely hydrodynamic configuration
  (see bottom left panel in Fig.~\ref{fig:initial}).
\end{itemize}

\begin{table}[h]
  \caption{\label{tab:table1} Summary of the initial model
    parameters. Here ${\mathcal{M}}/{T}$ is the ratio of the magnetic to
    the rotational kinetic energy, $\bar{B}$ is the magnetic-field
    strength computed via Eq.~(\ref{eq:B_field}) and $\beta^{-1}_{\rm
      ext}$ is the magnetic to gas pressure ratio in the stellar
    exterior. }
  \begin{ruledtabular}
    \begin{tabular}{cccc}
      Case&${\mathcal{M}}/{T}$ & $\bar{B}\times({M}/{10^6\,M_\odot})$G&$\beta^{-1}_{\rm ext}$\\
      \hline
     $\rm S_{\rm{\tiny{Int+Ext}}}$ &0.1& $6.5\times 10^6$ &100\\
     $\rm S_{\rm{\tiny{Int}}}$     &0.1& $6.5\times 10^6$ &0\\
     $\rm S_{\rm{\tiny{Hydro}}}$   &0  & 0   &0\\
    \end{tabular}
  \end{ruledtabular}
\end{table}

The magnetic field in the magnetized configurations we consider is
generated by the two-component vector potential
\begin{align}
  A_\phi =e^{-(r/r_1)^{2p}}A^{(1)}_\phi+
  \left(1-e^{-(r/r_1)^{2p}}\right)\,A^{(2)}_\phi\,,
  \label{eq:A_full}
\end{align}
where $r^2 = (x-x_{\text{\tiny{SMS}}})^2+(y-y_{\text{\tiny
    SMS}})^2+z^2$ with ($x_{\text{\tiny SMS}}$, $y_{\text{\tiny
    SMS}})$ the coordinates of the center of mass of the star. The
constants $r_1$ and $p$ are free parameters that control the radial
position and the width of the transition region between the two vector
potentials $A^{(1)}_\phi$ and $A^{(2)}_\phi$.  The vector
potential $A^{(1)}_\phi$ is given by
\begin{align}
  A^{(1)}_\phi = A_{ b}\,\varpi^2\,\text{max}
  (P-P_{\text{cut}}, 0)^{n_b}\,,
  \label{eq:A_int}
\end{align}
with $\varpi^2 = (x-x_{\text{\tiny SMS}})^2+(y-y_{\text{\tiny
    SMS}})^2$, $P_{\rm cut}$ the cutoff pressure that confines the
magnetic field to a region where $P>P_{\rm cut}$, and $A_{ b}$ the
constant that adjusts the initial magnetic-field strength. Here $A^{1}
_\phi$ is used for seeding a poloidal magnetic field for the $S_{\rm
  Int}$ case; i.e., effectively we set $r_1=\infty$ in
Eq.~\eqref{eq:A_full}. Vector potentials of this type with $n_b=1$
have been used for studying magnetized accretion disks around
stationary black holes~\cite{McK20004,VilHawKro03} and in compact
binary mergers involving neutron stars (see, e.g. Ref. ~\cite{Etienne:2012te}
and references therein), but here we set $n_b=1/8$ to approximate the
interior magnetic-field configuration that was adopted
in~\cite{LiuShaSte07}. For the case $\rm S_{\rm{\tiny{Int}}}$ we set
$P_{\rm cut}=10^{-4}P_{\rm max}$, with $P_{\rm max}$ being the maximum
value of the pressure at $t=0$. For the case $\rm S_{\rm{\tiny{Int}}}$ we
use a standard constant-density atmosphere with rest-mass density
$\rho_{0,\,{\rm atm}}=10^{-10}\,\rho_{0,\,\rm max}$, where
$\rho_{0,\,\rm max}$ is the maximum value of the rest-mass density at
$t=0$.

The vector potential $A^{(2)}_\phi$ is given
by~\cite{Paschalidis:2013jsa}
\begin{align}
A^{(2)}_{ \phi} = \frac{\pi\,r_0^2\,I_0\,\varpi^2}{(r^2_0+r^2)^{3/2}}\,
\left(1+\frac{15\,r_0^2\,(r^2_0+\varpi^2)}{8(r^2_0+r^2)^2}\right)\,,
  \label{eq:A_ext}
\end{align}
and approximates the magnetic field generated by a current loop, which
becomes a dipole at large distances. Here $r_0$ and $I_0$ are the loop
radius and current, and they determine the geometry and strength of
the magnetic field.

For the $\rm S_{ Int+Ext}$ case we use the superposition of the two
vector potentials because $A^{(2)} _{ \phi}$ alone does not appear to
have enough degrees of freedom to allow us to specify both the total
magnetic energy and the value of the plasma parameter $\beta$ in the
stellar exterior as we discuss below. The form~\eqref{eq:A_full}
guarantees a rapid and smooth transition of the magnetic field from
$A^{(1)} _{\phi}$ in the stellar interior to $A^{(2)} _{\phi}$ in the
exterior (see top left panel in Fig.~\ref{fig:Evolution}).  We adopt
$P_{\rm cut}=10^{-4}P_{\rm max}$, $r_0 \approx 2.2 M$, $r_1\approx 240
M$, and $p = 2$. Although this choice of superposed vector potentials
does not necessarily correspond to a realistic distribution of
currents, it allows a fairer comparison with the interior-only case because the bulk of the interior magnetic field in the $S_{\rm
  Int+Ext}$ case is practically the same as in the $S_{\rm Int}$ case,
and any differences arise because of the exterior
component. Following~\cite{PasRuiSha15}, to mimic a force-free
magnetosphere in the stellar exterior and to reliably evolve the
magnetic field outside the star in the $S_{\text{Int+Ext}}$ case, at
$t=0$ we set a low and variable density atmosphere in the exterior
such that the magnetic to gas pressure ratio is $\beta^{-1}_{\rm
  ext}=100$. 

We set $A_{b} =  2.91 \times 10^{-7}$, $I_0 = 7.35\times10^{-3}$ for
the $\rm S_{Int+Ext}$ case, and $A_{b} =  1.26 \times 10^{-6}$,
$I_0 = 2.25\times10^{-3}$ for the $\rm S_{Int}$ case. These values
fix the ratio of magnetic to rotational kinetic energy $\mathcal{M}/T$
to be 0.1 (corresponding to $\mathcal{M}/|W| = 9 \times 10^{-4}$).
Hence, the magnetic field is dynamically unimportant initially. We compute
the magnetic energy as measured by a normal observer $\mathcal{M}$ through
Eq. (30) of~\cite{LiuShaSte07}. Table~\ref{tab:table1} summarizes the
initial parameters of our models.

The resulting averaged magnetic-field strength $\bar{B}$ is
\begin{align}
  \bar{B}\equiv \sqrt{8\,\pi\,\mathcal{M}/V_s}=
  6.5\times 10^6\,\left(\frac{M}{10^6\,M_\odot}\right)\rm G\,,
  \label{eq:B_field}
\end{align}
which matches the initial averaged magnetic-field strength used
in~\cite{LiuShaSte07}. Here $V_s=\int \sqrt{\gamma}\,d^3x$ is the
proper volume of the star at $t=0$, and $\gamma$ is the determinant of
the three-metric $\gamma_{ij}$.

To accelerate the collapse, at $t=0$ the pressure is depleted by $1\%$
for all three cases.

\subsection{Evolution}

We use the Illinois GRMHD adaptive mesh refinement (AMR)
code embedded in  the Cactus/Carpet infrastructure
\cite{cactus_web,carpet_web}. Note that this code is different
than its publicly available counterpart embedded in the Einstein
Toolkit~\cite{Etienne:2015cea}.
This code has been widely tested and
used in different scenarios involving compact objects and/or magnetic
fields~(see,
e.g. Refs. ~\cite{Etienne:2007jg,lset08,Paschalidis:2010dh,PasLiuEti11,
  Etienne:2012te,Gold:2013zma,Gold2014}). For
implementation details see
\cite{Etienne:2010ui,Etienne:2011ea,EtVpas12}.

The Illinois code solves the equations of ideal GRMHD in a flux
conservative formulation [see Eqs. (27)--(29) in~\cite{Etienne:2010ui}]
via high-resolution shock capturing methods \cite{DueLiuSha05}. To
guarantee that the magnetic field remains divergenceless, the code
solves the magnetic induction equation via a vector potential
formulation [see Eqs. (8) and (9) in \cite{EtVpas12}]. We adopt the
generalized Lorenz gauge~\cite{EtVpas12,FarGolPas12} to close
Maxwell's equations, and employ a damping parameter $\xi=4.6/M$, where
$M$ the ADM mass of the system. This EM gauge choice avoids the
development of spurious magnetic fields that arise due to
interpolations across AMR levels (see~\cite{EtVpas12} for more
details).

The GRMHD evolution equations are closed by employing a $\Gamma$-law
EOS, $P=(\Gamma-1)\,\epsilon\,\rho_0$, which allows for shock
heating. Here $\epsilon$ is the specific internal energy and $\rho_0$
the rest-mass density. In all our models we set $\Gamma=4/3$, which is
appropriate when thermal radiation pressure dominates~\cite{BauSha99}.

To evolve the spacetime metric, we use the
Baumgarte-Shapiro-Shibata-Nakamura formulation of Einstein's
equations~\cite{ShiNak95, BauSha98b} coupled to the moving puncture
gauge conditions~\cite{BakCenCho05,CamLouMar05} with the equation for
the shift vector cast in first-order form (see,
e.g. Refs. ~\cite{HinBuoBoy13,RuiHilBer10}). The shift vector parameter $\eta$
is set to~$\eta = 4.6/M$.

\subsection{Grid structure}
During collapse, the equatorial radius of the star shrinks from $\sim
630M$ to a few $M$. To follow the evolution efficiently we add
high-resolution refinement levels as the collapse proceeds. This same
approach was also adopted in~\cite{ShiSha02,LiuShaSte07}.  In all the
cases listed in Table~\ref{tab:table1}, we begin the numerical
integrations by using a set of five nested refinement levels differing
in size and resolution by factors of 2. The base level has a
half-side length of $1312 M \approx 2.1\,R_{\rm eq}\approx 1.9\times
10^{9}(M/{10^6 M_{\odot}})$km, which sets the location of the outer
boundary. The grid spacing on the base level is $21.8M=3.22 \times
10^{7}(M/{10^6 M_{\odot}})$km. To save computational resources,
reflection symmetry across the equatorial plane is imposed. The
resulting number of grid points per level is $N=120^2\times 60$. To
maintain high resolution throughout the collapse, we add a new
refinement level with the same number of grid points $N$, and half the
grid spacing of the previous highest-resolution level every time the
density increases by a factor of 3. Such a procedure is repeated 5
times for the $\rm S_{\rm{\tiny{Hydro}}}$ case and 6 times for the $\rm
S_{\rm{\tiny{Int}}}$ and $\rm S_{\rm{\tiny{Int+Ext}}}$ cases.

Thus, in the last stages of the collapse the grid structure consists
of a total of eleven (ten) nested refinement levels in the MHD
(hydrodynamic) evolutions, in which the finest level has grid spacing
of $\sim 0.021M=3.1\times 10^{4}(M/{10^6 M_{\odot}})$km [$\sim
0.042M=6.2\times 10^{4}(M/{10^6 M_{\odot}})$km]. The highest
resolution on our grids is similar to that used in the axisymmetric
simulations of~\cite{LiuShaSte07}, but now our simulations are in 3+1
dimensions. The main purpose of applying higher resolution in the
magnetized cases is to more accurately evolve the low-density, near
force-free environments that emerge above the black hole poles.

\subsection{Diagnostics}
As a check on the validity of the numerical integration we monitor
the Hamiltonian and momentum constraints computed in Eqs.(40) and (41)
in~\cite{Etienne:2007jg}.  In all our cases, the normalized constraint
violations remain below $1\%$ over the entire evolution. We also check
the conservation of the rest mass $M_0$, and monitor the
ADM mass $M$ and the ADM angular momentum $J$. These quantities  are
computed by performing the ADM mass and angular momentum integrals via
Eqs. (21) and (22) in~\cite{Etienne:2011ea} over the surface of coordinate
spheres.
A fraction of the system's mass and angular momentum are radiated away
through gravitational and EM radiation as well as by escaping matter.
The dominant loss through our outermost extraction sphere is via escaping
matter (see Table ~\ref{tab:table2}), but that corresponds
to only $1\%$ of the ADM mass by the end of our
simulations. Therefore, we are reassured to find that the ADM
mass is conserved to $\sim 1\%$, and that the ADM angular momentum
is conserved to $\sim 2\%$ in all of our cases.

We use our modified version of the Psikadelia thorn to extract
GWs through the Weyl scalar $\Psi_4$, which is decomposed into $s=-2$
spin-weighted spherical harmonics (see, e.g. Refs. ~\cite{RuiAlcNun07}).
We estimate the thermal energy generated by shocks through the entropy
parameter $K\equiv P/P_{0}$, where $P_{0}=\kappa\,\rho_0^\Gamma$ is the
pressure associated with the
unshocked EOS. The specific internal energy has a ``cold''
$\epsilon_0$ and a ``thermal'' component $\epsilon_{\rm th}$, i.e.,
$\epsilon=\epsilon_{0} +\epsilon_{\rm th}$
with~\cite{EtiLiuSha08}
\begin{align}
\epsilon_{0}&= -\int P_{0}\,d(1/\rho_0)=
 \frac{\kappa}{\Gamma - 1}\,\rho_0^{\Gamma-1}\,.
\label{eq:Ecold}
\end{align}
Using the $\Gamma$-law EOS, it is straightforward to show that
$\epsilon_{\rm th}=(K-1)\,\epsilon_{0}$. Thus, for shock-heated
gas ($\epsilon_{\rm th}>0$) the entropy parameter always
satisfies $K>1$~\cite{EtiLiuSha08}.

We adopt the AHFinderDirect thorn~\cite{Tho03c} to locate the
apparent horizon (AH) following BH formation, and we use the isolated horizon
formalism to estimate the dimensionless spin parameter $a_{\rm
  BH}/M_{\rm BH}$ and mass $M_{\rm BH}$ of the BH~\cite{DreKriSho02}.

Finally, following BH formation the outgoing EM luminosity is computed
as in~\cite{Paschalidis:2013jsa,Ruiz:2012te} through the following
surface integral:
\begin{align}
  L_{\rm EM}=
  \lim_{r\rightarrow\infty}\int r^2\,S^{\hat r}\,d\Omega\,.
\label{eq:Lem}
\end{align}
The surfaces of integration are spheres of constant coordinate radii
at large distances from the BH.
Here $S^{\hat{r}}$ is the Poynting vector
$\mathbf{S}=(\mathbf{E}\times \mathbf{B})/4\pi$ projected onto the
outgoing unit vector $\hat{r}$.


\begin{table*}[t]
  \caption{\label{tab:table2} Summary of key results.  Here $t_{\rm BH}$
    is the coordinate time at which the apparent horizon
    appears, $M_{\rm BH}$ and $a_{\rm BH}/M_{\rm BH}$ are the mass and
    dimensionless spin parameter of the BH after they settle down (at
    $t-t_{\rm BH}\sim 150 M$), $M_{\rm esc}$ is the rest mass of
    unbound matter, $M_{\rm disk}$ is the rest mass outside the
    horizon minus the unbound mass, and $\dot{M}$ is the rest-mass
    accretion rate. The last two quantities have been computed after
    the accretion rate has settled, and $\tau_{\rm disk}=M_{\rm
      disk}/\dot{M}$ is the disk lifetime. The quantities $t_{\rm
      BH}$ and $\tau_{\rm disk}$ are normalized by $(M/10^6M_\odot)$, and
    $\Gamma_L$ is an average Lorenz factor within the funnel. For the
    $\rm S_{\rm{\tiny{Int+Ext}}}$ case $\Gamma_L$ is quoted at the
    time when the ratio $b^2/(8\pi\,\rho_0) \sim 200$ above the BH
    poles, where $b$ is the magnetic-field strength measured by an
    observer comoving with the plasma. For the $\rm
    S_{\rm{\tiny{Int}}}$ case $\Gamma_L$ is quoted near the end of the
    simulation, $L_{\rm EM}$ is the time-averaged Poynting luminosity
    over the last $300M$ before we terminate our simulations, and
    $L_{\rm GW}$ is the time-averaged GW luminosity over the duration
    of the GW burst $\Delta t_{\rm GW}\simeq 80M$. In the
    magnetized cases, the anticipated total energy removed by EM
    processes, $E_{\rm EM} \sim L_{\rm EM}\times \tau_{\rm disk} \sim
    10^{-5}-10^{-3} M$, exceeds the total energy lost in GWs $E_{\rm GW}\simeq
    10^{-6}M$.}
  \begin{ruledtabular}
    \begin{tabular}{cccccccccccccc}
      Model                  &$t_{\rm BH}$ & $M_{\rm BH}/M$  &  $a/M_{\rm BH}$ & $M_{\rm esc}/M$ &$M_{\rm disk}/M$&$\dot{M}(M_\odot/s)$&$\tau_{\rm disk}$ &$\Gamma_L$&$L_{\rm EM}$ erg/s        &$L_{\rm GW}$ erg/s \\
      \hline
      $\rm S_{\rm{\tiny{Int+Ext}}}$ &$1.5\times 10^5$s  & 0.91        &   0.71      &  $1.1\%$        &  $7.0\%$     &  1.11                &$7.2\times 10^4$s& 1.20     &$10^{52.5}$           &$4.7\times 10^{51}$\\
      $\rm S_{\rm{\tiny{Int}}}$     &$1.5\times 10^5$s  & 0.92        &   0.75      &  $0.9\%$        &  $6.0\%$     &  1.20                &$5.0\times 10^4$s& 1.20     &$10^{50.6}$           &$4.7\times 10^{51}$\\
      $\rm S_{\rm{\tiny{Hydro}}}$   &$1.4\times 10^5$s  & 0.92        &   0.75      &  $0.2\%$        &  $9.0\%$     &  1.0                 &$9.0\times 10^4$s &-         &-                     &$4.7\times 10^{51}$\\
    \end{tabular}
  \end{ruledtabular}
\end{table*}

\section{Results}
\label{section:Numerical Results}
The early stages of the evolution are similar for all three cases we
consider. Thus, we focus the discussion on the $\rm
S_{\rm{\tiny{Int+Ext}}}$ case, unless otherwise specified. Key results
from all cases are summarized in Table~\ref{tab:table2}.

%
\begin{figure}
  \includegraphics[scale=0.13]{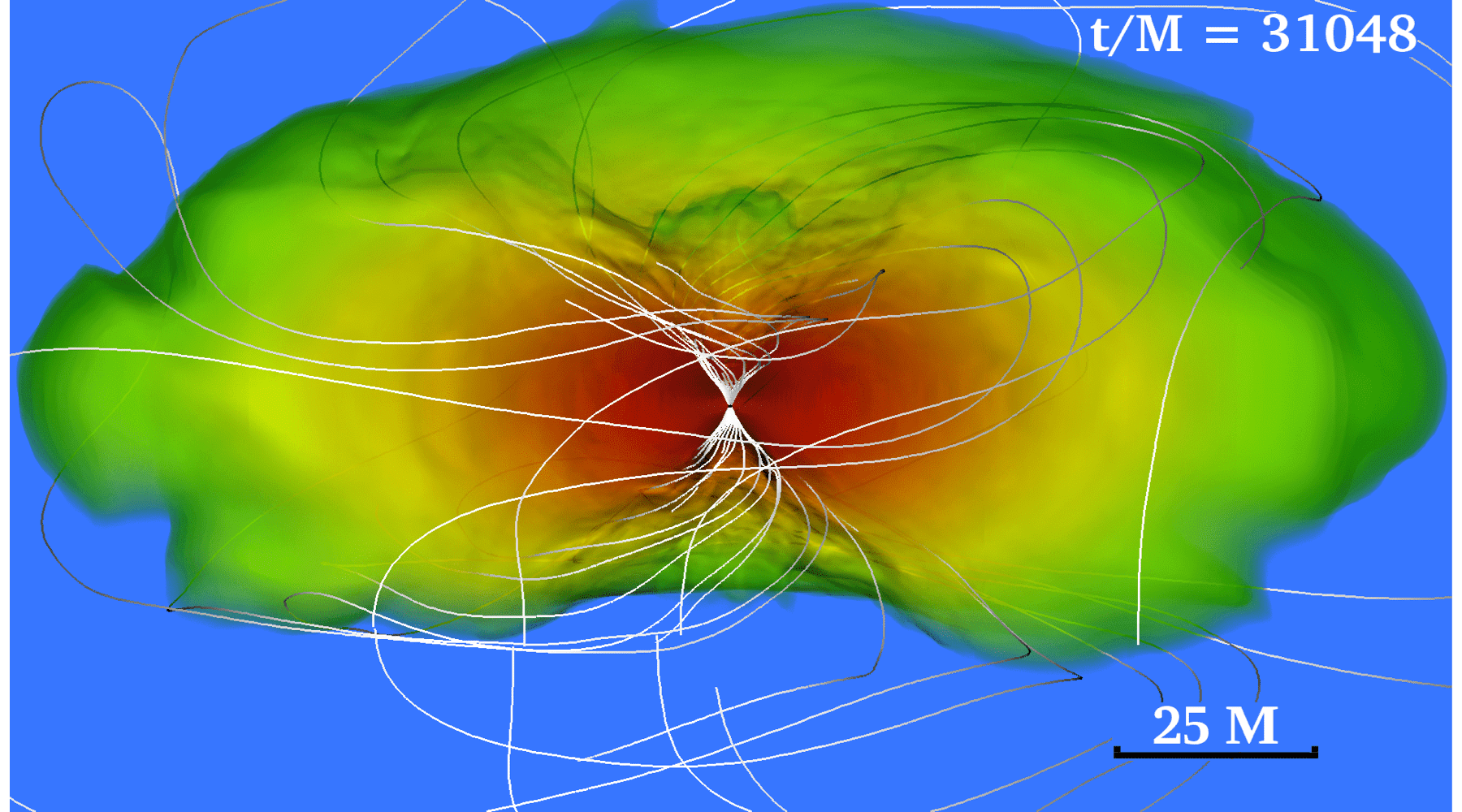}
  \includegraphics[scale=0.13]{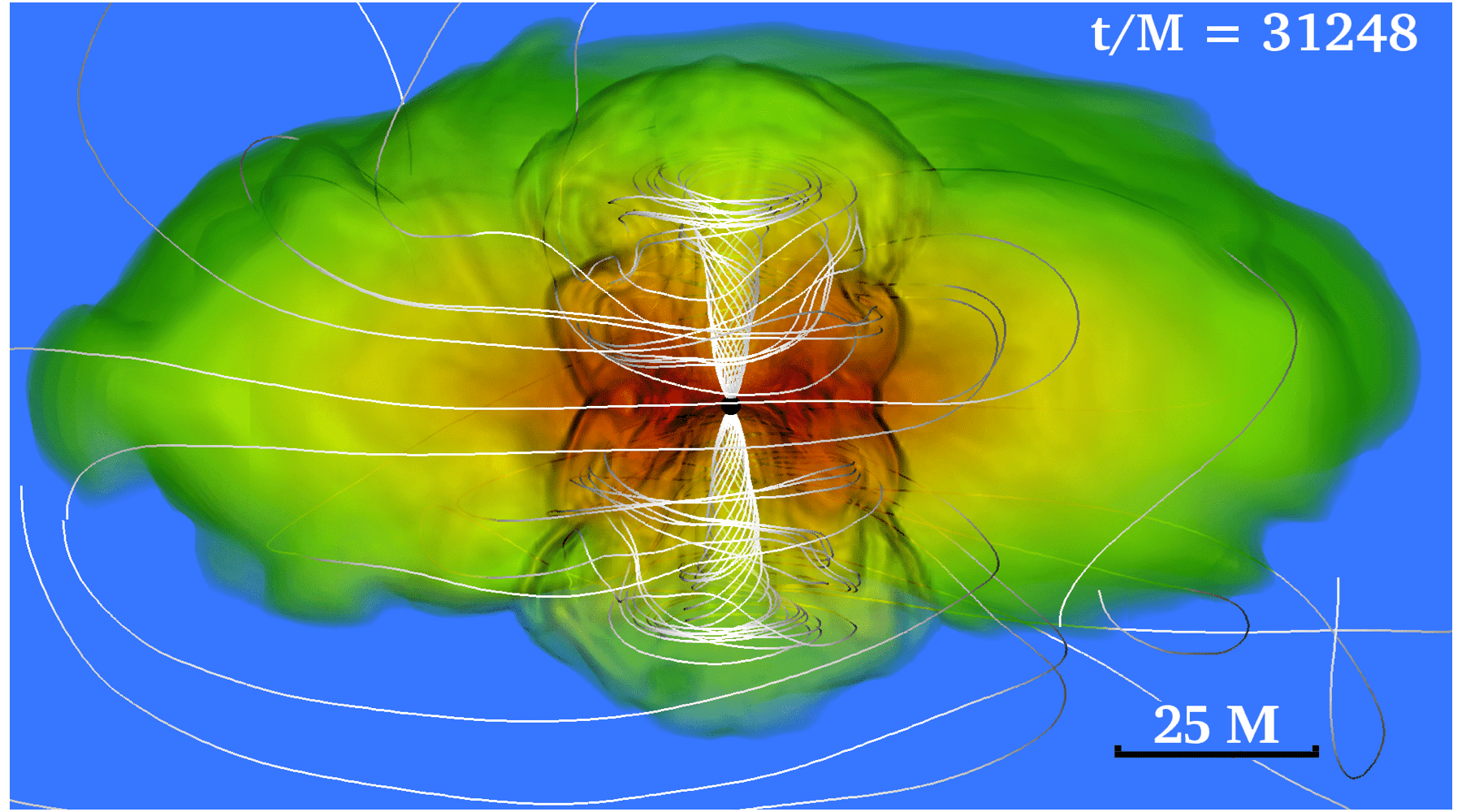}
  \includegraphics[scale=0.13]{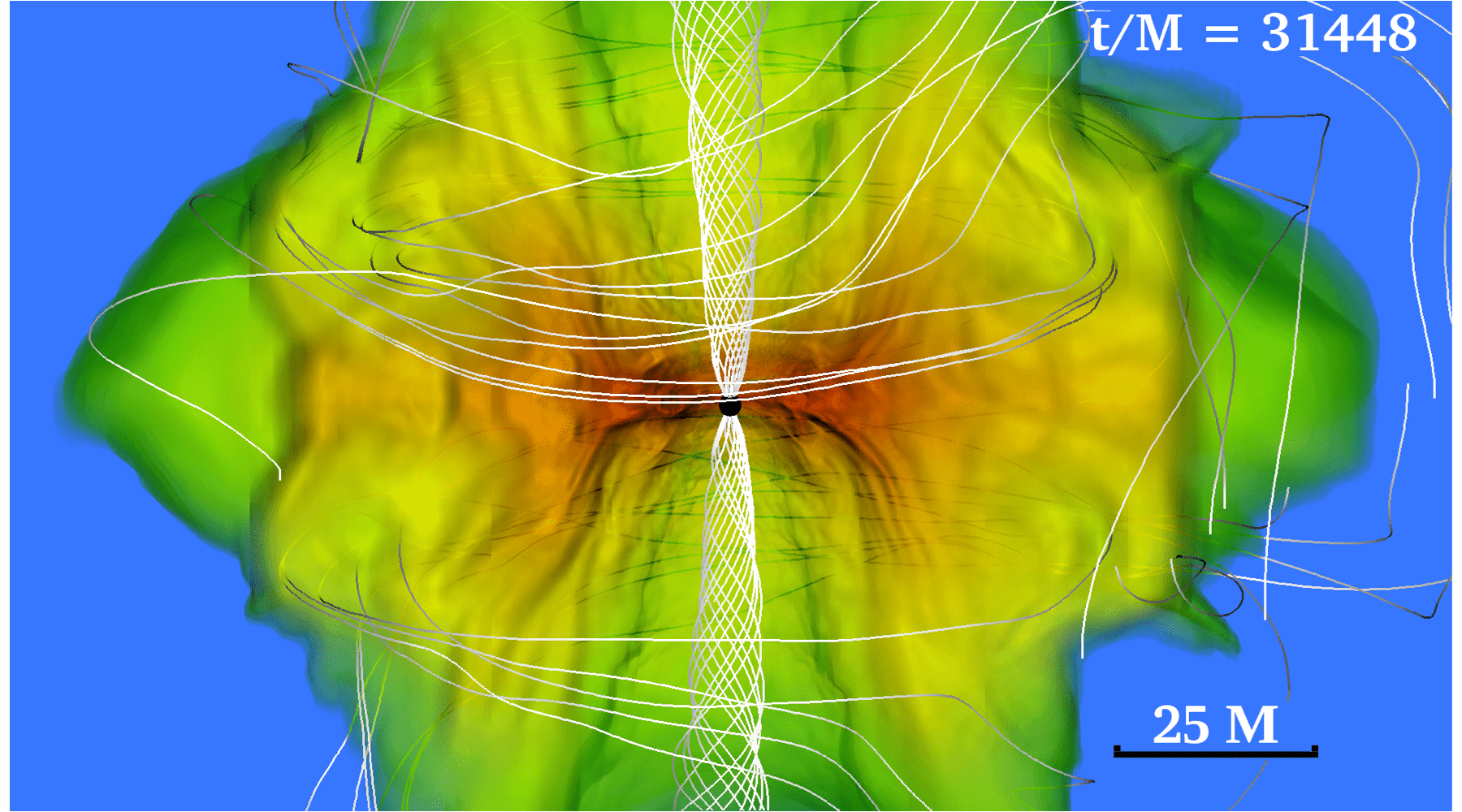}
  \caption{\label{fig:K} Meridional cut of 3D density profile for the $\rm
    S_{\rm{\tiny{Int+Ext}}}$ case at $t-t_{\rm{BH}} = \text{ 0,
      200}M\text{, and 400} M$, with magnetic-field lines in white. The
    top panel corresponds to the time near BH formation. The middle
    panel shows the shock front propagating outward along which
    the entropy parameter $K$ becomes $>1$. The shock drives an
    outflow which eventually becomes a magnetically supported and confined
    incipient jet (bottom panel).}
\end{figure}

Following the initial pressure depletion the star undergoes collapse
(top right panel in Fig.~\ref{fig:Evolution}). As the gas falls
inward, the density in the stellar interior increases. By about
$t\sim1.3\times 10^4M\simeq 6.4\times 10^4 (M/10^6M_\odot){\rm s}$ we
observe the formation of an inner core that undergoes rapid
collapse. Similar behavior was found in the Newtonian simulations of a
$\Gamma=4/3$ polytrope in~\cite{GolWeb80}. In addition to the
increasing matter density, we observe that during the last stages of
the collapse the frozen-in magnetic-field lines are compressed and
become wound (middle, left panel in Fig.~\ref{fig:Evolution}), and
the magnetic energy builds up rapidly and is amplified by a factor of
$\sim 100$ until a BH forms. During this period, we resolve the
wavelength of the fastest-growing magnetorotational-instability (MRI)
mode by $\gtrsim 10$ points---the rule-of-thumb for capturing
MRI~\cite{ShiLiuSha06}. MRI acts as an effective viscosity driving
turbulence and thus helps maintain the accretion of gas onto the
BH once the system reaches quasistationary equilibrium. In the early
stages immediately following collapse, however, hydrodynamical forces
drive the accretion, and the rate for the pure hydrodynamical and
magnetic-field cases are comparable. MRI also contributes to the
amplification of the poloidal magnetic field, while magnetic winding
amplifies the toroidal component. This amplification occurs both in
the disk and above the BH poles.

The AH appears approximately at the same time
$t_{\rm BH}$ in all cases, which is expected because the seed magnetic
field is dynamically unimportant initially. Right after the AH
appearance, the mass and spin of the remnant BH evolve rapidly as the
surrounding gas is accreted. Following this high-accretion episode,
the rapid growth of the BH settles at about $t-t_{\rm BH}\sim
150M\approx 740(M/10^6M_\odot)$s. At this time the values of the BH
mass and dimensionless spin are $M_{\rm BH}\approx 0.91M$ and $a_{\rm
  BH}/M_{\rm BH}\approx 0.71$ for the $\rm S_{\rm{\tiny{Int+Ext}}}$
case, and $M_{\rm BH}\approx 0.92M$ and $a_{\rm BH}/M_{\rm BH}\approx
0.75$ for the other two cases (see Table~\ref{tab:table2}).  These
values are consistent with those of the previous axisymmetric calculations
of~\cite{ShiSha02,LiuShaSte07}.

Following BH formation, high-angular-momentum gas originating in the
outer layers of the star begins to settle in an accretion torus around
the BH (see middle left panel
in Fig~\ref{fig:Evolution}). During this
phase, a substantial amount of gas descends towards the BH, which
increases the density in the torus. The rapidly swirling, dense gas
soon forms a centrifugal barrier onto which additional infalling
matter collides, and ultimately a reverse shock is launched at
$t-t_{BH}\sim 170M=830(M/10^6M_\odot)$s (see Fig.~\ref{fig:K}). The
shock increases the entropy of the gas and pushes the fluid outward.
This initial outflow
ultimately turns into a wind  which is almost
isotropic. The entropy parameter $K$ exceeds 1 in all three cases.

In the hydrodynamic case the shock-driven, isotropic outflow
disappears after $t\sim 1100M$ (see bottom right panel in
Fig.~\ref{fig:initial}). By contrast, in the magnetized cases the
initial outflow develops into one with two components: an isotropic,
pressure-dominated wind component, and a collimated, mildly
relativistic, Poynting-dominated component---an incipient jet. In
particular, the magnetic-field lines anchored into the initial
shock-driven outflow are stretched, forming a poloidal component, and
they become more tightly wound (see middle right panel in
Fig~\ref{fig:Evolution}). Magnetic winding converts poloidal to
toroidal flux and builds up magnetic pressure above the BH poles in a
similar fashion as discussed in~\cite{PasRuiSha15} for black
hole--neutron star mergers. Eventually, the growing magnetic pressure
gradients become so strong that an outflow is launched and sustained
by the helical magnetic fields. During this period, the magnetic field
above the BH pole reaches a value of $\sim 4.0\times
10^{10}({10^{6}M_{\odot}/M})$G and remains roughly constant.

As the magnetic pressure above the BH poles increases for $t > t_{\rm
  BH}$, magnetically dominated regions where $b^2/(8\pi\rho_0)>1$
(where $b$ is the magnetic-field strength measured by an observer
comoving with the plasma) expand outwards above the BH poles, forming
an incipient jet (see collimated, helical magnetic field in the bottom
left panel in Fig~\ref{fig:Evolution} and top right panel in
Fig.~\ref{fig:initial}). Based on the distribution of the outgoing
flux on the surface of the distant sphere we estimate that the half-opening
angle of the jet is~$\sim 25^\circ$. We define the jet half-opening angle, as the polar angle $\theta_0$ at which the Poynting
flux drops to $50\%$ of the maximum. In contrast to the hydrodynamic
case, in the magnetized cases the outflow persists until we terminate
our simulations because it is driven by the magnetic field.

\begin{figure}
  \includegraphics[scale=0.13]{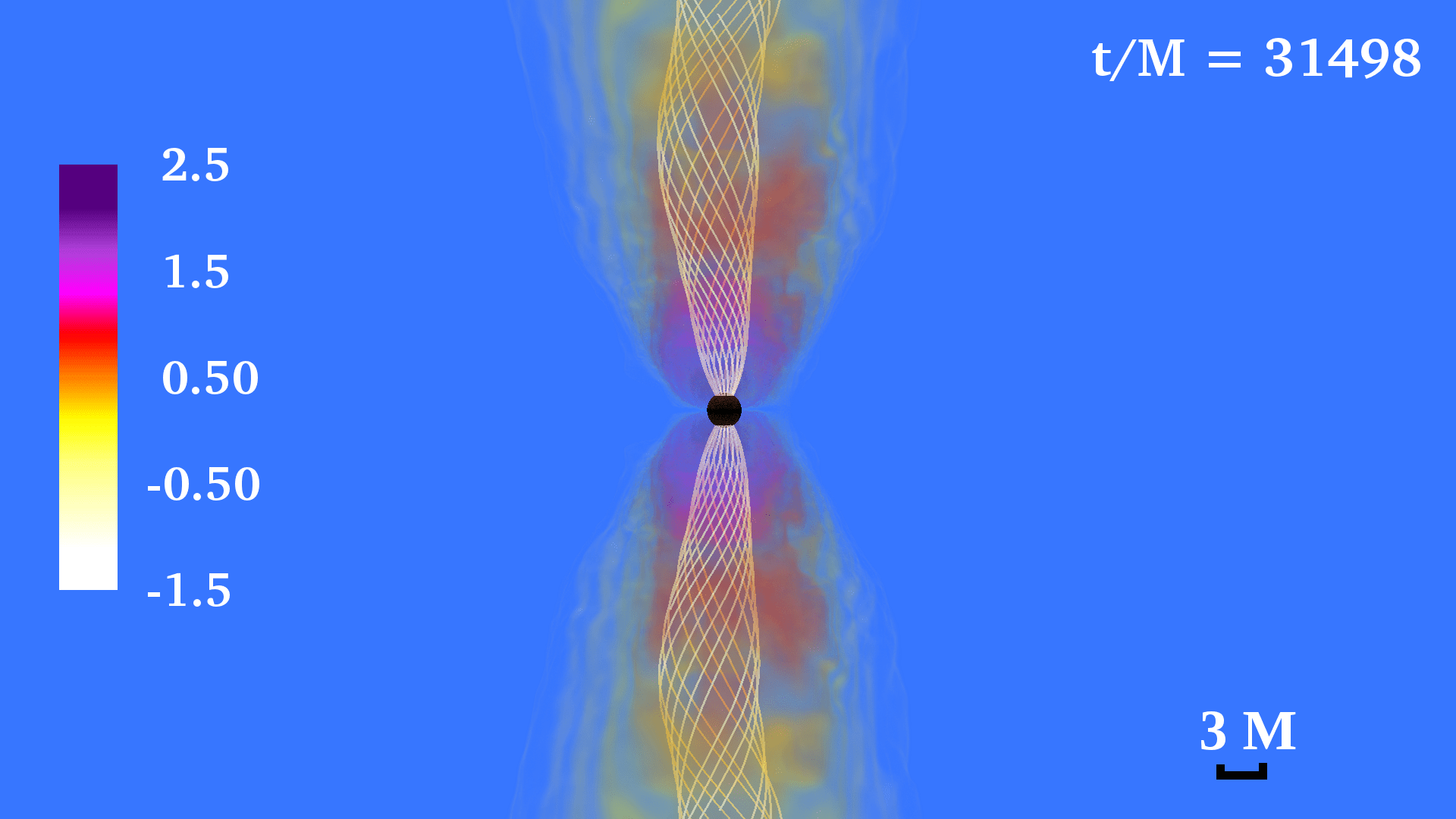}
  \caption{\label{fig:b2rho} Ratio of magnetic energy density to
    rest-mass density $b^2$/$(8\pi\rho_0)$ (log scale) at $t-t_{\rm
      BH}\approx 450M$ for the $\rm S_{\rm{\tiny{Int+Ext}}}$ case. The
    helical magnetic-field lines (solid curves) are plotted in the collimated
    funnel with $b^2/(8\pi\rho_0) \geq
    10^{-1.5}$. Magnetically dominated areas ($b^2/8\pi\rho_0 \geq 1$)
    extend to heights greater than $50M\approx 50\,r_{\rm BH}$ above the
    BH horizon (black sphere).}
\end{figure}

The characteristic value of the Lorentz factor measured by a normal
observer at large distances ($\Gamma_L=\alpha\,u^0$, with $\alpha$
being the lapse function) in the funnel is $\Gamma_L \approx 1.2$.
The outflow is therefore mildly relativistic. However, the value of
the magnetization in the funnel becomes $b^2/(8\pi\rho_0)\gtrsim
100$. This is shown in Fig.~\ref{fig:b2rho} which displays a volume
rendering of the magnetization at $t-t_{\rm BH}\approx
450M\approx 2200(M/10^6M_\odot)$s. Highly magnetized regions extend to
$\gtrsim 50M\approx 50 r_{\rm BH}$ above the BH poles (here $r_{\rm
  BH}$ is the apparent horizon radius). The ratio $b^2/(8\pi\rho_0)$
equals the terminal Lorentz factor in axisymmetric, steady-state,
magnetically dominated jets~\cite{Vlahakis03}. Thus, the incipient
jets found here, in principle, can be accelerated to typical Lorentz
factors required by GRB observations~\cite{Gehrels:2013xd}. However,
the terminal Lorentz factor is anticipated to be reached at hundreds
of thousands to millions of $M$ away from the
engine~\cite{TchMckNar07, Paschalidis16} outside of our computational
domain. We note that although our code may not be reliable at values
of $b^2/(8\pi\rho_0)\gtrsim200$, the increase in the magnetization in the
funnel is robust (see discussions
in~\cite{PasRuiSha15,RuiLanPas16}). As
in~\cite{PasRuiSha15,RuiLanPas16}, to ensure the physical nature of
the jet, we track Lagrangian particle tracers and ensure that the
matter in the jet is being replenished by plasma originating in the
torus and not in the artificial atmosphere.

\begin{figure}[h]
  \hspace*{-0.35cm}\includegraphics[scale=0.17]{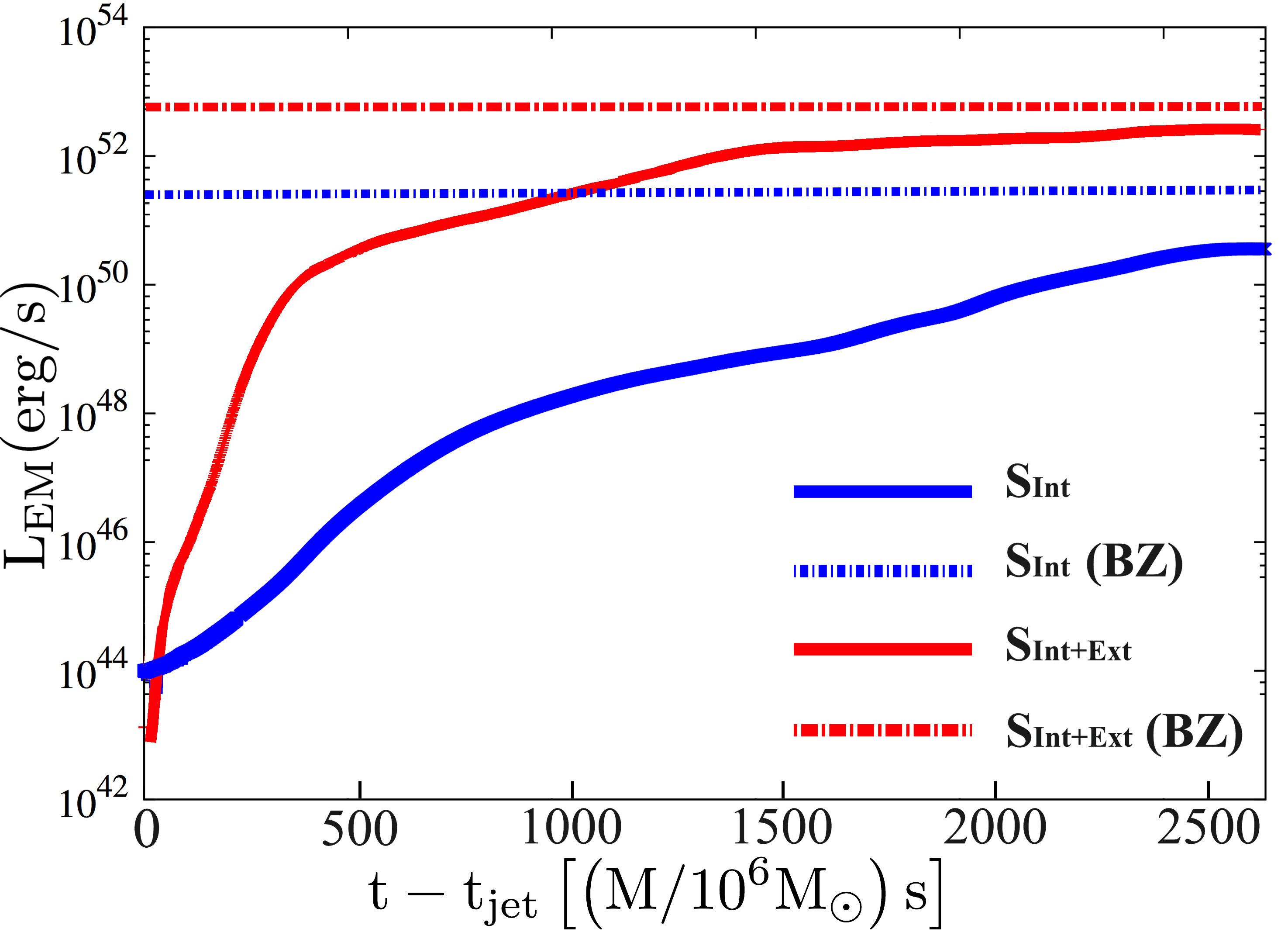}
  \caption{\label{fig:Lem} Poynting luminosity $L_{\rm EM}$ vs time
    $t\geq t_{\rm{jet}}$ calculated on a sphere with coordinate radius
    $175M=2.43\times 10^8(M/10^6M_\odot)\rm km$ for
    the magnetized cases as displayed in Table~\ref{tab:table2}
    (continuous lines). Here $t_{\rm jet}$ defines the time when the jet front
    reaches 100M above the BH ~\cite{RuiLanPas16}. The dashed part in
    the $\rm S_{\rm Int+Ext}$ curve
    indicates the region where the ratio $b^2/(2\,\rho_0)$ becomes
    $\gtrsim 200$. In that region, our numerical results may not be
    reliable.  The dotted-dashed lines show the expected BZ luminosity
    computed via Eq.~(\ref{eq:LumBZ}) for $M_{\rm BH}$. }
\end{figure}

In all three cases outgoing matter ($v^r = u^r/u^0 > 0$) in the jet
funnel and wind, which reaches distances $r\gtrsim 100M\simeq 1.47\times
10^8(M/10^6M_\odot)\rm km$, becomes unbound ($E=-u_0-1>0$). The mass
fraction ($M_{\text{esc}}/M$) ejected in the $\rm
S_{\rm{\tiny{Hydro}}}$, $\rm S_{\rm{\tiny{Int}}}$, and
$S_{\rm{\tiny{Int+Ext}}}$ cases is $\sim 0.2\%$, $0.9\%$, and $1.1\%$,
respectively (see Table~\ref{tab:table2}). The values of the unbound
mass in the $S_{\rm{\tiny{Hydro}}}$ and $\rm S_{\rm{\tiny{Int}}}$
cases are in close agreement with the values reported
in~\cite{LiuShaSte07}.  These results demonstrate that the magnetic
fields enhance the amount of unbound mass, a result which is also
consistent with the fact that we observe jets in both magnetized
cases. Figure ~\ref{fig:Lem} shows $L_{\rm EM}$ as a function of time for
the two magnetized cases, where we see that it is
$10^{51}-10^{52}\rm erg\ s^{-1}$. This luminosity is comparable to
those we found for black hole--neutron star~\cite{PasRuiSha15} and
neutron star--neutron star~\cite{RuiLanPas16} mergers, quite different
scenarios. This implies that there is enough energy to power a typical
GRB in all of these events~\cite{shapiro2017}. This luminosity 
implies that BH--disks formed following the collapse of either SMSs or
massive Pop III stars can power GRBs. Notice that the luminosity is larger
in $\rm S_{Int+Ext}$ than that in the $\rm S_{Int}$ case. There are a
few differences between the $\rm S_{Int+Ext}$ model and the $\rm S_{Int}$
model that can explain this effect. First, the very outer layers of the SMS
are magnetized in the $\rm S_{Int+Ext}$ model but not in the $\rm S_{Int}$
model. Note that it is these very outer layers which form the outer
layers in the remnant disk, from which fluid particles escape and go
into the jet funnel. Second, the exterior in the $\rm S_{Int+Ext}$
mimics a force-free environment, while in the $\rm S_{Int}$ case it does not
(there is no magnetic field in the exterior). Thus, it is easier to
``punch" a hole in the exterior in the $\rm S_{Int+Ext}$ model than in the
$\rm S_{Int}$ model because of less baryon loading. These differences are
likely the source of the differences in the jet power observed. Also notice
that, unlike ~\cite{McKBla09, MotRicOtt14}, there are no prominent kink
instabilities present in our simulation. During the whole evolution our
disk remains roughly axisymmetric and is not characterized by any
significant $m=1$ density perturbation.

%

To determine if the magnetized outflow is powered by the
Blandford-Znajek (BZ) process~\cite{BlaZna77}, we compare the EM
luminosity computed via Eq.~(\ref{eq:Lem}) with the following analytic
BZ estimate~\cite{BlaZna77,ThoPriMac86}:
\begin{equation}
  \begin{multlined}
    L_{BZ}\approx 10^{51}
    \left(\frac{a_{\rm BH}/M_{\rm BH}}{0.75}\right)^2\, \left(\frac{M_{\rm BH}}
         {10^6M_{\odot}}\right)^2
    \left(\frac{B_{\rm BH}}{10^{10}\text{G}}\right)^2 \text{erg/s}\,,
  \end{multlined}
  \label{eq:LumBZ}
\end{equation}
and show the result in Fig.~\ref{fig:Lem}.

Note that in this expression for $B_{\rm BH}$, we use the time-averaged
value of the magnetic field that is measured by a normal observer over
the last $300M$ before we terminate our simulations. Here $B_{\rm BH}$
scales like $1/M$. Given that $M_{\rm BH}$ scales like $M$, the actual
parameter fixed by our simulations is $M_{\rm BH}B_{\rm BH}$: by fixing
$\mathcal{M}/|W|$ ($\sim M^2B^2$ in geometrized units), we fixed the
product $M B$. In other words, for our collapse scenario, both the product
$M_{\rm BH}B_{\rm BH}$ and hence $L_{BZ}$ are independent of the initial
$M$~\cite{shapiro2017}. We find that $L_{\rm EM}~\sim10^{52.5}\rm
erg\ s^{-1}$ on an extraction sphere with coordinate radius
$175M=2.43\times 10^8(M/10^6M_\odot)\rm km$; thus, this is consistent with
the BZ process. In addition, we check the ratio of the angular
frequency of the magnetic-field lines to the black hole angular
frequency, which is expected to be $\Omega_{F}/\Omega_{H}=0.5$ for a
split-monopole force-free magnetic-field
configuration~\cite{McK20004}. Here $\Omega_F = F_{t\theta}/
F_{\theta\phi}$ is the angular frequency of magnetic field, with
$F_{\mu\nu}$ the Faraday tensor, and the angular frequency of the
black hole is defined as ~\cite{Alc08}
\begin{align}
  \Omega_H = \frac{(a_{\rm BH}/M_{\rm BH})}{2M_{\rm BH}}\,
  \left(1+\sqrt{1-(a_{\rm BH}/M_{\rm BH})^2}\right)\,.
\end{align}  
We compute this ratio in magnetically dominated regions on an
azimuthal plane passing through the BH centroid and along coordinate
semicircles of radii $1.05\,r_{\rm BH}\leq r\leq 1.5 r_{\rm BH}$. We
find that, within an opening angle of $\theta \sim 20^\circ-30^\circ$
from the black hole rotation axis, $\Omega_{F}/\Omega_{H}\approx
0.2-0.35$. As it has been pointed out
in~\cite{PasRuiSha15,RuiLanPas16}, the deviation from the value 0.5
could be due to the deviations from strict stationarity and
axisymmetry of the spacetime, the non-split-monopole geometry of the
magnetic field in our simulations, the gauge used to compute
$\Omega_F$, and/or insufficient resolution. Despite this discrepancy,
the results suggest that the BZ effect is likely operating in our
simulations.

\begin{figure}[h]
  \hspace*{-0.4cm}\includegraphics[scale=0.067]{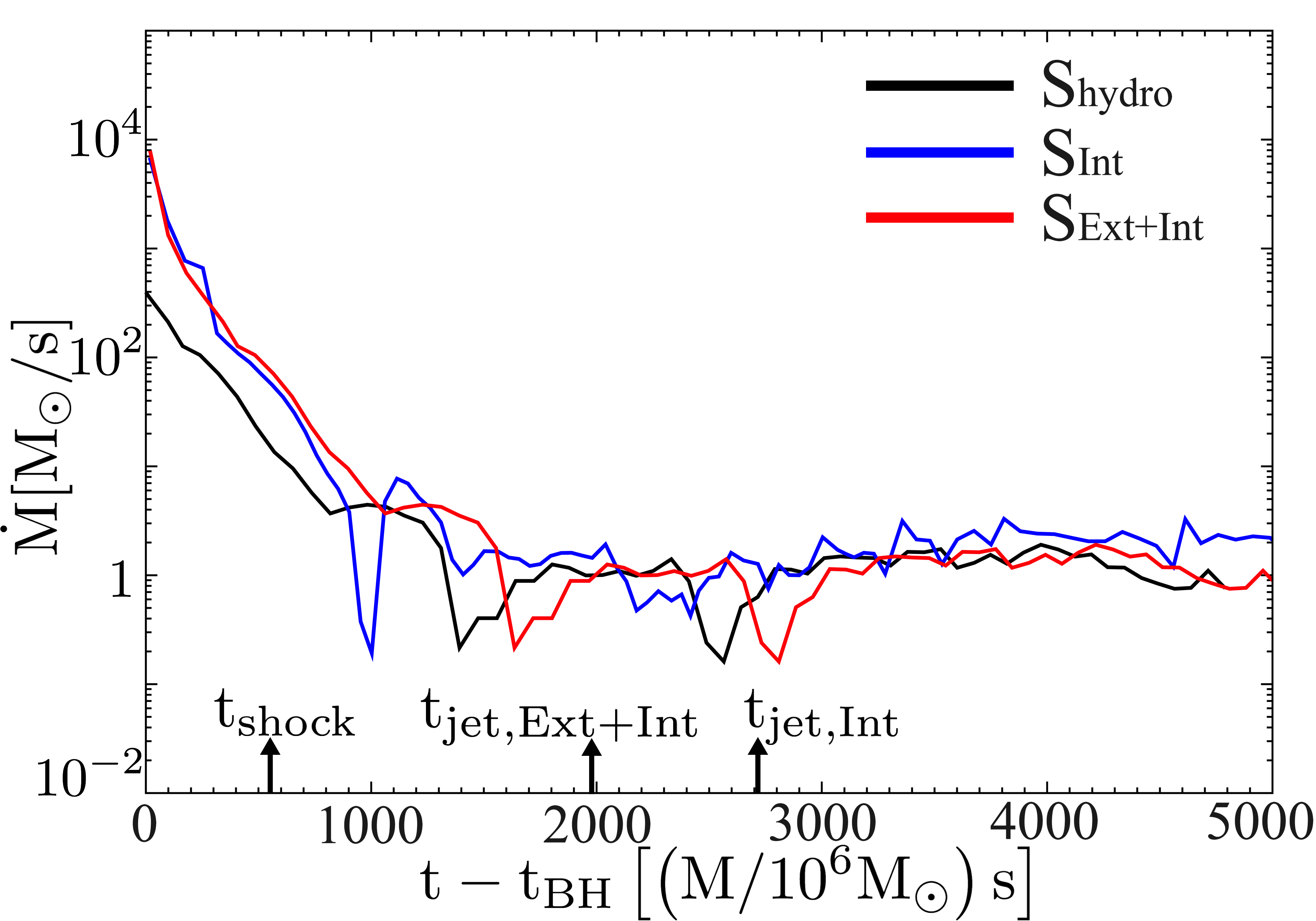}
  \caption{\label{fig:Mdot} Rest-mass accretion rate $\dot{M}$ for all
    the cases listed in Table~\ref{tab:table1}. The arrows denote the
    shock wave formation and the jet launching times, which are
    defined in the same way as in~\cite{PasRuiSha15}.}
\end{figure}

As displayed in Fig~\ref{fig:Mdot}, the accretion rate settles to
$\dot{M}=1.1\,M_\odot/$s by $t-t_{\rm BH}\approx 370M=1.8\times
10^3(M/10^6M_\odot)$s, at which time the mass of the accretion torus
is $M_{\rm disk}=7\times 10^4\,M_\odot(M/10^6M_\odot)$.  Thus, the
duration of the jet which is fueled by the torus is expected to last
for an accretion time $\Delta t=M_{\rm disk}/\dot{M}\sim 6\times
10^4(M/10^6M_\odot)$s, which is consistent with the estimates
in~\cite{MatNakIok15}. Combining this result with the outgoing
Poynting luminosity, we find that the amount of energy anticipated to
be removed via electromagnetic processes after an accretion time scale
is $\sim 10^{-5}-10^{-3} M$. By contrast, the amount of energy lost in
GWs is $E_{\rm GW}\simeq 10^{-6}M$ (see Table~\ref{tab:table2}). Thus,
our simulations indicate that collapsing SMSs with mass $\sim
10^4-10^5M_\odot$ are viable jet engines for ultra-long GRBs such as
the 25000s-long GRB 111209A~\cite{Gendre2013ApJ} (though it is not
likely that GRB 111209A is related to SMSs since it is observed at a
redshift of $z=0.68$), while those with mass larger than $10^6M_\odot$
do not seem to fit within the GRB phenomenon. On the other hand, our
results indicate that collapsing Pop III stars with mass $M \gtrsim
240M_\odot$, are viable engines for long GRBs.

\section{Observational Prospects}
\label{section:Observational Prospects}
Detection of an EM signal coincident with a GW would mark  a
``golden moment'' in multimessenger astronomy.
A simultaneous detection of GW and EM signals
with the signatures summarized below would provide direct evidence
for the existence of SMSs, and hence provide a major
breakthrough in understanding the cosmological formation of SMBHs.
In the following section, we discuss the prospects for detecting
multimessenger signatures of collapsing SMSs.

\subsection{Gravitational waves}
\label{subsection:Gravitational Waves}
\begin{figure}[h]
  \hspace*{-0.25cm}\includegraphics[scale=0.175]{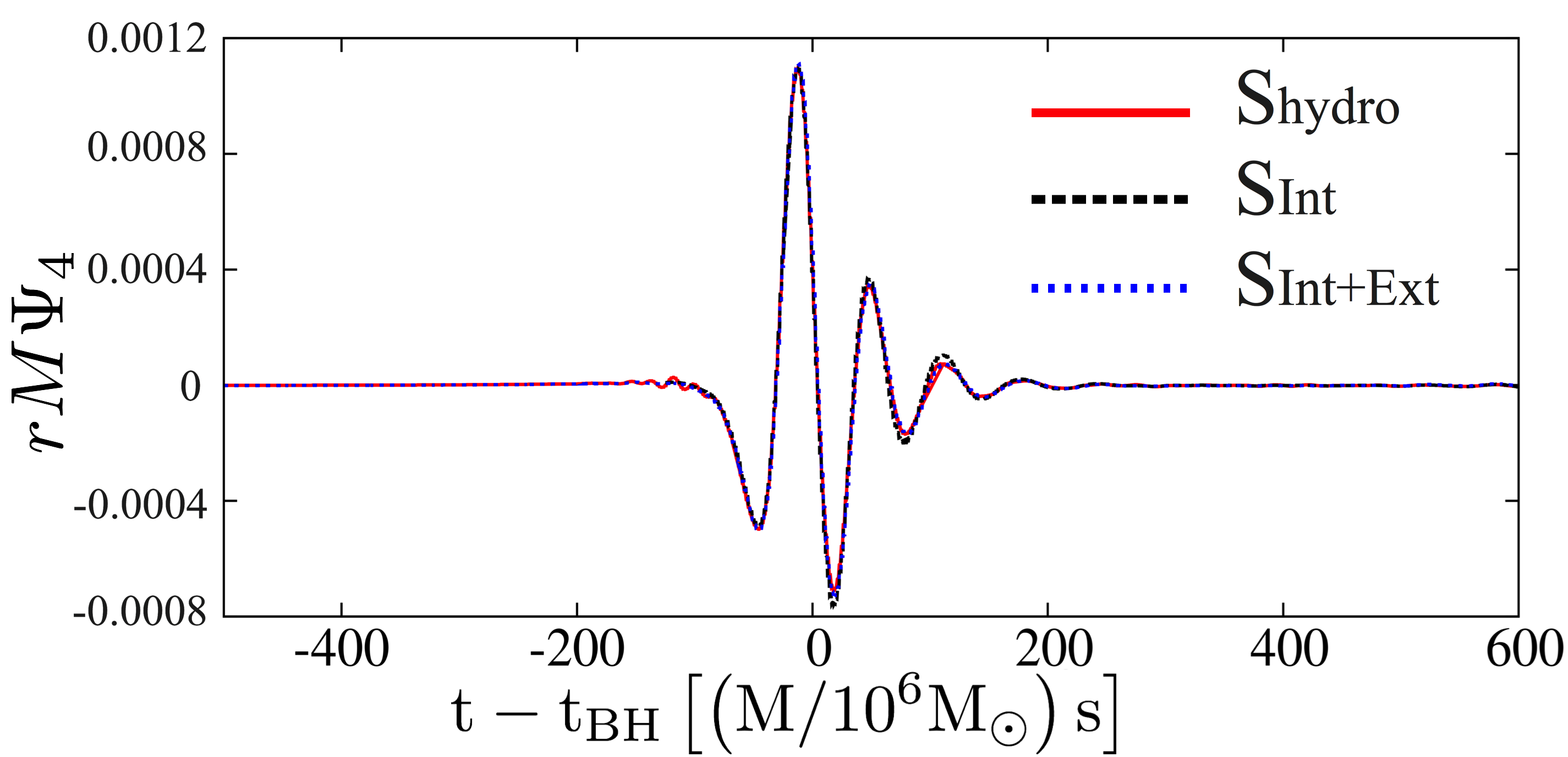}
  \caption{\label{fig:psi4} Real part of the $(l,m)=(2,0)$ mode of
    $\Psi_4$ versus time. We have shifted the time in all cases by the
    coordinate time of black hole formation.}
\end{figure}

In Fig.~\ref{fig:psi4} we plot the evolution of the real part of the
$(l,m)=(2,0)$ mode of $\Psi_4$. Given that the collapse proceeds almost
axisymmetrically, the $(l,m)=(2,0)$ mode is the dominant one. As the
figure demonstrates, there are no significant differences in the
waveform among the three cases we consider in this work. The amplitude
of $m\neq 0$ modes is smaller than $3\%$ of the $(l,m)=(2,0)$ mode,
demonstrating that deviations from nonaxisymmetry remain small
throughout the evolution. The oscillation period of the dominant mode
after BH formation
is $\sim 13M \approx 15.5M_{\rm BH}$, which corresponds to a frequency
of $f \approx 15.6(10^{6}M_{\odot}/M) /(1+z){\rm mHz}$. This
value is close to the expected quasinormal mode frequency of the
$(l,m)=(2,0)$ Kerr mode~\cite{BerCarWil05}. We find that our waveforms are in
qualitative agreement with the one obtained from axisymmetric GR,
purely hydrodynamic simulations of a SMS which is modeled as a
$\Gamma=1.335$ polytrope in~\cite{ShiSekUch16}.
%

\begin{figure}[h]
  \hspace*{-0.305cm}\includegraphics[scale=0.1035]{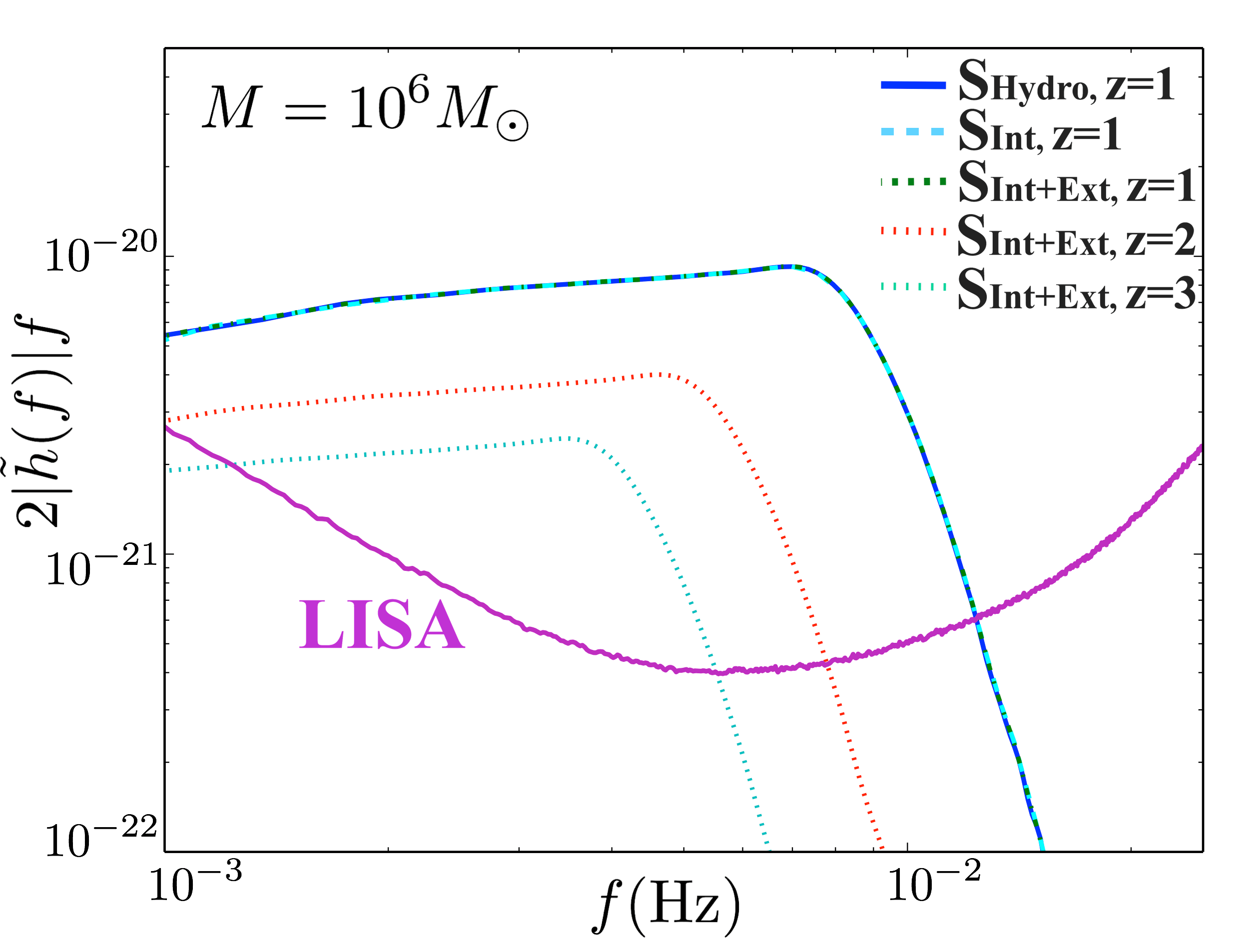}
  \hspace*{-0.38cm}\includegraphics[scale=0.151]{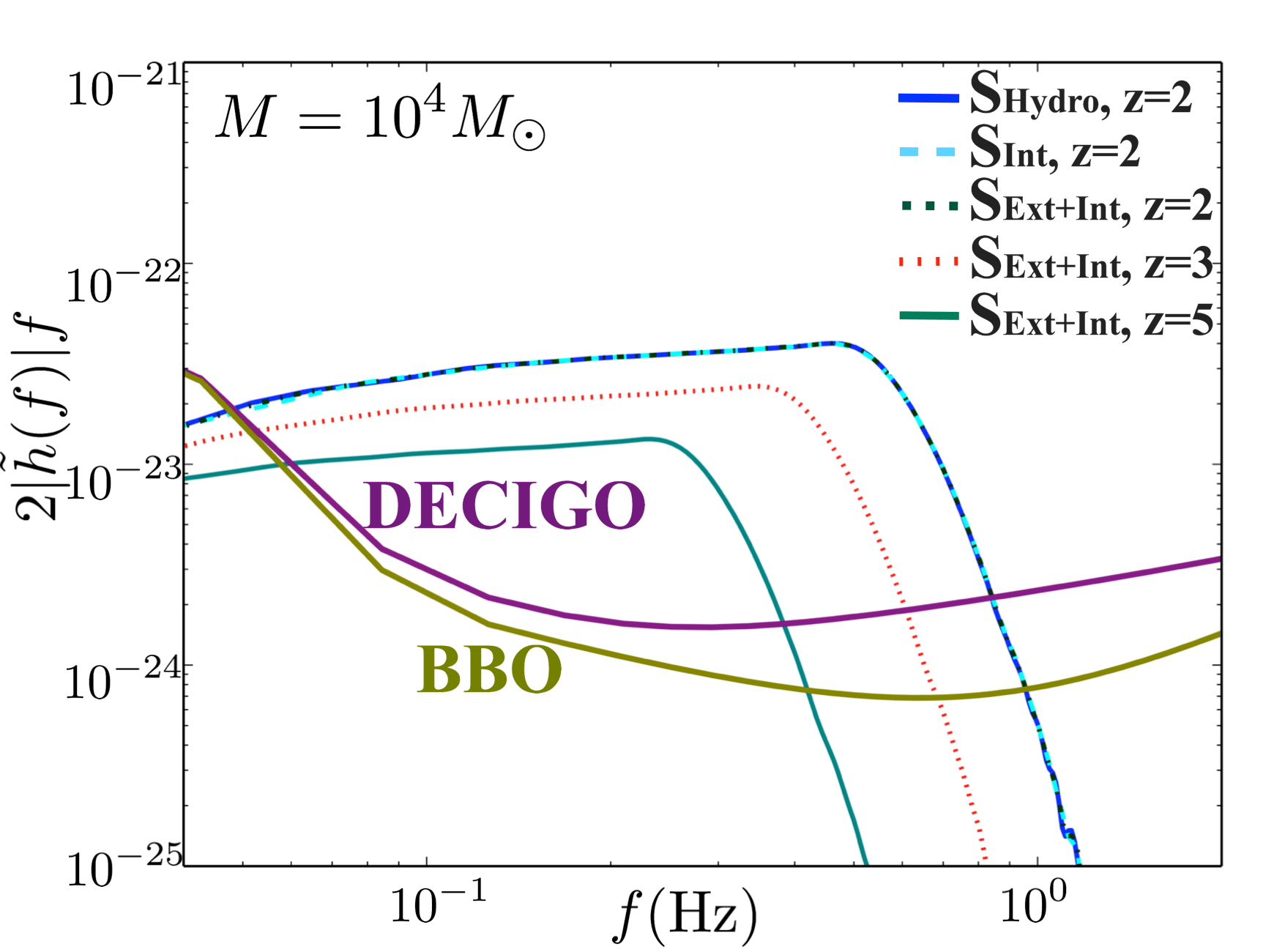}
  \caption{\label{fig:hf} For all of our models, we show $2|\tilde h(f)|f$ vs. frequency/ Top panel: The SMS mass is $M=10^{6}M_\odot$. The dashed
    and dotted curves denote the signal strength at redshift $z=1,2$,
    and 3. The solid curve corresponds to the LISA noise
    amplitude. Bottom panel: The SMS mass is $M=10^{4}M_\odot$. The
    dashed and dotted curves denote the signal strength at redshift
    $z=2$, 3, and 5. The solid curves correspond to the DECIGO and BBO
    noise amplitude as indicated in the plot. }
\end{figure}

To assess the detectability of GWs produced by SMS collapse, we
compute the strain amplitude $|\tilde h(f)| = \sqrt{\tilde h_+(f)^2
  +\tilde h_{\times}(f)^2}$ from $\Psi_4$ and compare it to the
expected LISA sensitivity curve~\cite{eLISA_sensitivity}.
Here $\tilde h_{\times,+}(f)$ are the Fourier transforms of $h_{
  \times,+}(t)$. The top panel in Fig.~\ref{fig:hf} shows a plot of
twice the characteristic strain $2|\tilde h(f)|f$ for all three cases
listed in Table~\ref{tab:table1}, assuming $M = 10^6 M_{\odot}$, and
cosmological redshift of $z=1$. As expected, all three agree well with
each other. We also plot the GW spectra for the S$_{\text{Int+Ext}}$
case at $z =2$ and $z=3$, assuming $M = 10^6 M_{\odot}$, as well as
the LISA noise amplitude $[S(f)f]^{1/2}$ assuming the
configuration with four laser links between three satellites, and arm
length $L=5\times10^6$km~\cite{KleBarSes16}, which has acceleration
noise similar to what was found by the LISA Pathfinder
experiment~\cite{ArmAud16}. The peak value of the doubled
characteristic strain ($h_{2c} = 2|\tilde h(f)f|$) after taking the $
\theta$-averaged value of the $_{-2}Y^2_0$ spherical harmonic is
\begin{equation}
  h_{2c} \approx 9.2 \times 10^{-21} \left(\frac{M}{10^6
    M_{\odot}}\right)^{-1}\, \left( \frac{6.8 \text{Gpc}}{D_L}\right)\, .
\end{equation}
A source at luminosity distance $D_L = 6.8$ Gpc lies at redshift $z=1$
in a flat $\Lambda$-CDM cosmology with $H_0 = 67.6 \text{km s}^{-1}
\text{Mpc}^{-1}$ and $\Omega_{M} = 0.311$~\cite{Wri06,
  Grieb16}. Figure.~\ref{fig:hf} shows that the GW signal frequency lies
in the most sensitive part of the LISA sensitivity curve. We compute the
signal-to-noise ratio (SNR),
\begin{align}
{\rm SNR}^2=\int_0^\infty \frac{(2|\tilde h(f)|)^2}{S_n(f)}\,df\,,  
\end{align}
with $S_n(f)$ the one-sided noise spectral density of the detector,
and we find that for an optimally oriented source at redshift $z=3$,
${\rm SNR} \sim 7.4 $ for the LISA sensitivity curve used in
Fig.~\ref{fig:hf}. Thus, if SMSs could form and collapse at redshifts
$z\lesssim 3.0$, \textit{LISA} could detect their GW signature. This
is consistent with the axisymmetric simulations of~\cite{ShiSekUch16}.

For less massive progenitors ($10^4M_\odot$), the characteristic
strain would peak at the decihertz range. Thus, these sources would be
targets for future instruments like BBO ~\cite{HarFriSha06} and DECIGO ~\cite{TakNak03}. Despite the decrease in the
amplitude of the GWs due to the lower mass, the superior sensitivity
of decihertz GW detectors [$[S(f)f]^{1/2} \sim 10^{-24}$ at $f \sim
0.1$ Hz] makes these systems detectable at very large redshifts. The
bottom panel in Fig.~\ref{fig:hf} shows a plot of the
$\theta$-averaged doubled characteristic strain $2|\tilde h(f)|f$ for
all three cases listed in Table~\ref{tab:table1}, assuming $M = 10^4
M_{\odot}$, and a cosmological redshift of $z=2$. We also plot the GW
spectra for the S$_{\text{Int+Ext}}$ case at $z =3$ and $z=5$,
assuming $M = 10^4 M_{\odot}$, and the DECIGO/BBO noise
amplitudes based on the analytic fits of
~\cite{Yagi:2009zz,Yagi:2011yu}, which account for foreground and
background noise sources in addition to the instrument
noise. Employing the same detector noise amplitude, we computed the
SNR for $M=10^4M_\odot$, and found that optimally oriented sources
could be detected by DECIGO at redshift $z\lesssim 8$ with
${\rm SNR}\gtrsim 8$, and by BBO at $z\lesssim 11$ with ${\rm
  SNR}\gtrsim 8$. Thus, if the rate of collapsing SMSs at high
redshifts is sufficiently high, the exquisite sensitivity of
DECIGO/BBO could provide smoking-gun evidence for the
existence of such stars and the formation of massive seed BHs.

%
\subsection{Electromagnetic signatures}
\label{subsection:Electromagnetic Radiations}

%
\begin{figure}[h]
   \hspace*{-0.40cm}\includegraphics[scale=0.265]{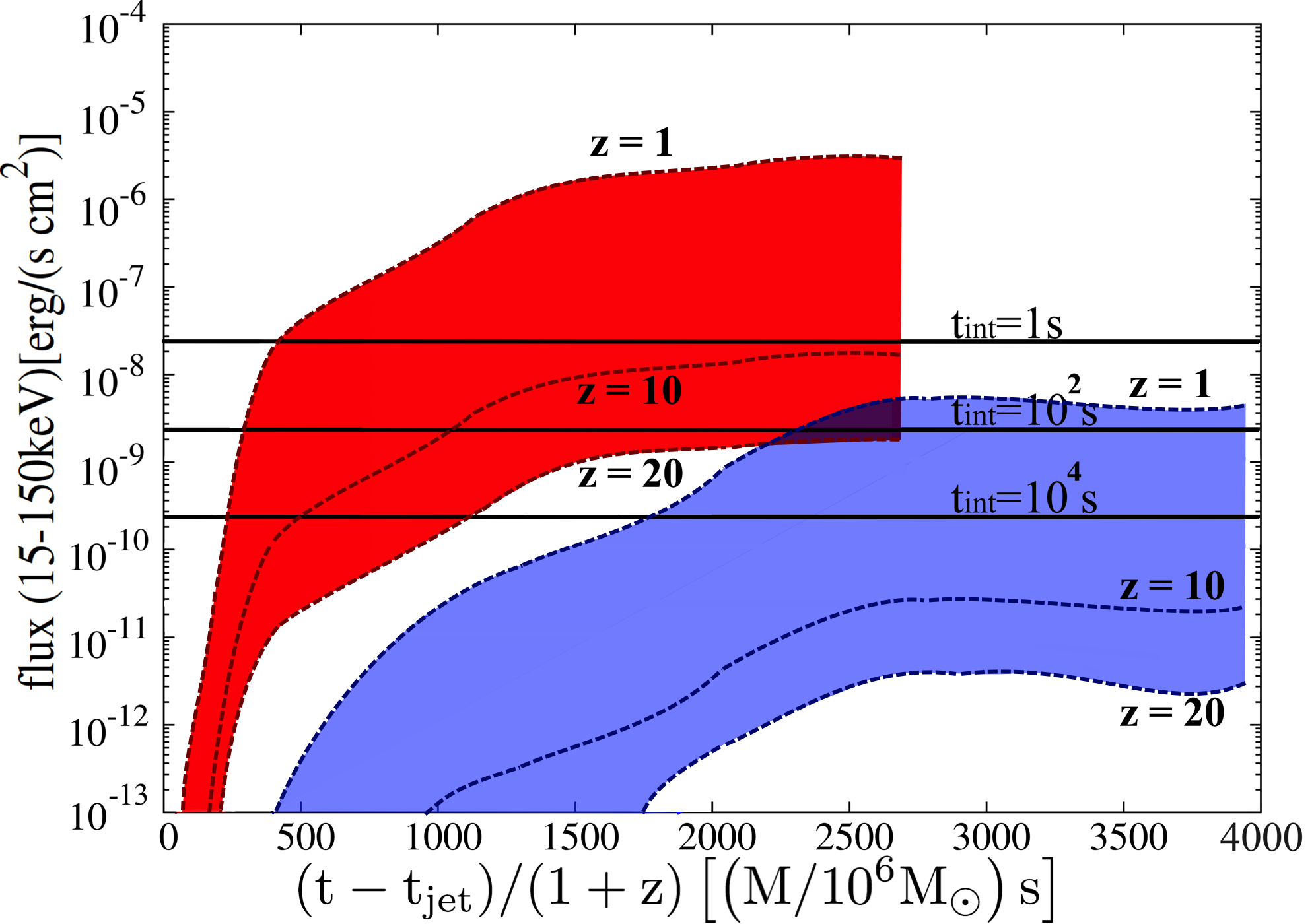}
  \caption{\label{fig:Lem_detect}  Gamma-ray flux versus time in the
    energy range $15-150$ keV at different redshifts for the $\rm
    S_{\rm{\tiny{Int+Ext}}}$ (top-red region) and $\rm
    S_{\rm{\tiny{Int}}}$ (bottom-blue region) cases.  The three
    horizontal lines show the sensitivity of BAT with integration time
    of $1$s, $10^2$s, and $10^4$s, from top to bottom.}
\end{figure}

To assess the detectability of the EM radiation from our magnetized
models by detectors such as the Swift's BAT, we assume that the following collapse a GRB-like event takes
place. We then compute the energy flux within BAT's energy range
(15--150 keV) in the observer frame as follows
\begin{equation}\label{eq:flux}
f = \frac{\epsilon L_{\rm EM}}{2 \pi \eta_{c} D^2_L(z)}\frac{\int^{150(1+z)\text{keV}}_{15(1+z)\text{keV}} E N(E) \text{d}E}{\int^{\infty}_{0}E N(E) \text{d}E},
\end{equation}
where $\epsilon$ is the fraction of the Poynting luminosity that becomes
photons, $\eta_{c}$ is a ``collimation'' factor, which
equals $2$ for isotropic emission and $0.2$ for a half-opening angle
of $25^\circ$, $D_L(z)$ is the luminosity distance, $N(E)$ is the
photon number spectral density in the source frame, and $L_{\rm{EM}}$ is
the outgoing Poynting luminosity we compute in our simulations.
Photons with energies in the range $15\textendash 150$ keV in the
observer frame, originate with energies $15(1+z)\textendash 150 (1+z)$
keV in the source frame.
Here, we approximate $N(E)$ by the
``GRB model" proposed in~\cite{BanMatFor93}, which consists of a
power-law continuum with an exponential cutoff at low energy that
continuously transitions to a steeper power law at high energy [see
Eq. (1) in ~\cite{BanMatFor93}]. In our calculation, the spectral
parameters $\alpha$, $\beta$, and $E_0$ of~\cite{BanMatFor93} are set
to -1, -2.3, and 150 keV, respectively~\cite{KanPreBri06}. In all our
estimates in this section, we also choose $\epsilon = 0.1$ and
$\eta_{c}=0.2$

In Fig.~\ref{fig:Lem_detect} we plot the total energy flux of
Eq.~\eqref{eq:flux} as a function of time for sources that are located
in the redshift range $1\leq z\leq 20$ and we compare it with BAT's
sensitivity at three different observation periods, $t_{\text{int}} =1,
10^2$, $10^4$s. The luminosity distance is computed assuming the
cosmological parameters we listed in the previous section. The
detector sensitivities for different observation periods are indicated
by the black horizontal lines. We estimate these using the BAT
sensitivity derived via a 70-month survey in the 14--195 keV
band~\cite{BayTueMar12}, and the fact that the sensitivity of BAT
approximately increases as $\sqrt{ t_{\rm{int}}}$, where $t_{\rm int}$ is
the integration or observation time (see, e.g., Ref.~\cite{BarBarCum05}).  We
find that for the $\rm S_{\rm{\tiny{Int+Ext}}}$ case, up to $z = 20$
the EM energy flux is greater than $10^{-10} \text{erg} \text{ /(s}
\text { cm}^2)$, which BAT is fully capable of detecting with
integration time $t_{\rm int} \sim 10^4$s. The results also hold
approximately for Fermi Gamma-ray Burst Monitor (GBM) whose sensitivity  is somewhat
smaller than Swift's ~\cite{BosGotBou14, ConPelBri13}. For the case
$\rm S_{\rm{\tiny{Int}}}$, a confident detection can be made up to
$z\sim 15$, but it would require integration time $t_{\text{int}}\sim
10^6$s, which may be too long. A characteristic-duration ULGRB may be
on the order of $10^4$s, which would require a disk lifetime of order
$10^3(1+z)$s in the SMS collapse scenario.
Such disk lifetimes could
arise for $M\sim 10^4 M_\odot$, for which we estimate that the ULGRB
detection could be made even at $z \sim 15$ in the $\rm
S_{\rm{\tiny{Int+Ext}}}$ scenario. Consequently, $M\sim 10^4 M_\odot$
SMSs that collapse at $z\sim 10$ are promising candidates for
coincident detection of multimessenger EM and GW signals. However, the
rates at which such events take place are uncertain and our results
motivate their study. Finally, our results suggest that collapsing Pop
III stars at redshift $z \sim 5-8$ could be the progenitors of long
GRBs that Swift and Fermi could detect. Hence, a
fraction of the high-redshift long GRBs that have already been
observed could have been powered by collapsing Pop III stars.

\section{Summary and Conclusions}
\label{section:Summary and Conclusions}
We performed magnetohydrodynamic simulations in 3+1 dimensions and
full general relativity of the magnetorotational collapse of $4/3$
polytropes, spinning initially at the mass-shedding limit and
marginally unstable. Our simulations model collapsing SMSs with masses
$\gtrsim 10^4M_\odot$, and they also crudely model collapsing, massive
Pop III stars. A major goal of our study was to assess the effects of
magnetic fields, and the multimessenger signatures of these
astrophysical objects. We extended previous studies by lifting the
assumption of axisymmetry and considered magnetic-field geometries
that are either completely confined to the stellar interior or extend
from the stellar interior out into the exterior.  We also considered a
purely hydrodynamic case in order to compare with previous GR
hydrodynamic simulations of SMS collapse (see,
e.g., Refs~\cite{ShiSha02,LiuShaSte07,Uchida:2017qwn}) and followed the
post-BH formation evolution for much longer times than previous
works. In our magnetized cases we ensured that the initial magnetic
field is dynamically unimportant by setting the ratio of the total
magnetic to kinetic energy to $0.1$, which corresponds to a
magnetic-to-gravitational-binding-energy ratio of
$\mathcal{M}/|W|=9\times 10^{-4}$.

In terms of the black hole mass, dimensionless black hole spin and
torus mass, the results from our hydrodynamic simulations are
consistent with previous semianalytic estimates and axisymmetric
simulation in GR reported
in~\cite{Sha04,ShiSha02,ShaShi02,LiuShaSte07}. We also find that
magnetic fields do not affect these global quantities~\cite{LiuShaSte07}.

In the magnetized cases, following BH formation, we observe the
formation of magnetically dominated regions above the black hole poles
where the magnetic-field lines have been wound into a collimated
helical funnel, within which the plasma flows outwards with a typical
Lorentz factor of $\Gamma_L\sim 1.2$. This collimated outflow is
mildly relativistic, and constitutes an incipient jet.  Our analysis
suggests that the Blandford-Znajek effect is likely operating in our
simulations and could be the process powering these jets. The
magnetization $b^2/(8\pi\,\rho_0)$ in the funnel reaches values
$\gtrsim 200$, and since for steady-state, axisymmetric jets the
magnetization approximately equals the jet terminal Lorenz factor, the
jets found in our simulations may reach Lorentz factors $\gtrsim 200$,
and hence explain GRB phenomena. The accretion torus lifetime is
$\Delta t\sim 10^5(1+z)(M/10^6M_\odot)$s. Thus, collapsing
supermassive stars with masses $10^3-10^4M_\odot$ at $z\sim 10-20$ are
candidates for ultra-long GRBs, while collapsing massive Pop III stars
at $z \sim 5-8$ are candidates for long GRBs. We estimated that for
observation times $\sim 10^4$s, Swift's BAT and Fermi's GBM could detect such ultra-long GRB events from
$10^3-10^4M_\odot$ supermassive stars at $z \lesssim 15$, and they could also
detect long GRB events from Pop III stars at $z \sim 5-8$. While
$10^6M_\odot$ supermassive stars could, in principle, power gamma-rays,
our models suggest that the burst duration at $z\sim 10$ would be
$10^6{\rm s} \sim 114 {\rm d}$ long, which would require long
integration times to observe.

Apart from sources of EM signals, we also demonstrated that
supermassive stars generate copious amounts of gravitational waves
with $(l,m)=(2,0)$ the dominant mode, and in agreement with the
axisymmetric results of~\cite{ShiSekUch16}. We find that if an
optimally oriented $10^6M_\odot$ SMS collapses to a BH at $z\lesssim
3$, its GW signature could be detectable by a LISA-like
detector, with a signal-to-noise ratio $\gtrsim 7.4$. Most
importantly, we point out that collapsing supermassive stars with
masses $10^4M_\odot$ generate gravitational waves which peak in the DECIGO/BBO bands, and that BBO (DECIGO)
could detect their GWs even at redshifts $z\lesssim 11$ ($z\lesssim
8$). Thus, we discover that $10^4M_\odot$ supermassive stars are
promising candidates for coincident multimessenger signals.

Some comments and caveats about our calculations are in order.  First,
our numerical results may not continue to be reliable for funnel
magnetizations $b^2/(8\pi\,\rho_0) \gtrsim 200$
(see, e.g. Ref. \cite{PasRuiSha15}), which is why we terminate our
simulations when such high values are reached.  However, based on
previous work and tests with our code we are confident that the
increase in the magnetization and jet launching is robust.  Moreover,
by the time we terminate, the BH-disk-jet configuration has settled
into quasistationary equilibrium even as the magnetization grows.
Second, we used a $\Gamma = 4/3$ $\Gamma$-law EOS to model our stars.
However, most observed long-gamma-ray bursts are believed to originate
from a Pop I star with an EOS that becomes stiffer once the core
density approaches nuclear density~\cite{MacWoo99}.  Third, we have
ignored pair-creation effects. Differential rotation may be present in
rapidly rotating stars, at least in outer layers~\cite{ShaTeu83}.
Hence an uniformly rotating model of a supermassive star may only be an
approximation. However, differential rotation may not be maintained in
turbulent magnetized scenarios~\cite{BauSha99}. We also neglect the
possibility of nuclear burning in the SMS core, although it is
unimportant for $M \gtrsim 10^5 M_\odot$~\cite{Uchida:2017qwn}.  We
also note that the collapse of differentially rotating supermassive
stars with small initial nonaxisymmetric density perturbations may
induce the formation of multiple black holes due to a fragmentation
instability, as has been reported in pure hydrodynamic studies
\cite{ZinSteHaw05,ZinSteHaw06,ReiOttAbd13}.
We plan to explore all of these aspects in future investigations.


\acknowledgements We thank the Illinois Relativity Group REU team
members Eric Connelly, Cunwei Fan, John Simone, and Patchara
WongSutthikoson for assistance in creating Figs.~\ref{fig:Evolution},
\ref{fig:initial}, and~\ref{fig:b2rho}.  We also thank Mitchell
Begelman and Kent Yagi for useful discussions. This work has been supported in part by National Science Foundation (NSF) Grants PHY-1602536 and PHY-1662211, and NASA Grants NNX13AH44G and 80NSSC17K0070 at the University of Illinois at Urbana-Champaign. V.P. gratefully acknowledges support from NSF Grants No. PHY-1607449, NASA grant NNX16AR67G (Fermi) and the Simons Foundation. This work used the
Extreme Science and Engineering Discovery Environment (XSEDE), which
is supported by NSF Grant No. OCI-1053575. This research is part of
the Blue Waters sustained-petascale computing project, which is
supported by the National Science Foundation (No. OCI
07-25070) and the state of Illinois. Blue Waters is a joint effort of
the University of Illinois at Urbana-Champaign and its National Center
for Supercomputing Applications.

\bibliographystyle{apsrev}
\bibliography{references}

\end{document}